\documentclass[10pt,aps,pra,twocolumn,superscriptaddress,floatfix,longbibliography]{revtex4-2}
\usepackage{placeins}
\usepackage[english]{babel}
\usepackage{xcolor}
\usepackage[utf8]{inputenc}
\usepackage{mathtools}
\usepackage{physics}
\usepackage{braket}
\usepackage{bbm}
\usepackage{xspace}
\usepackage{bm}
\usepackage{amsbsy}
\usepackage{amsthm}
\usepackage{amssymb}
\usepackage{lipsum}
\usepackage{amsfonts}
\usepackage{amsmath}
\usepackage{soul}
\setstcolor{red}
\usepackage{dsfont}
\usepackage{graphicx}
\usepackage{float}
\usepackage{hhline}
\usepackage{epsfig}
\usepackage{epstopdf}
\usepackage{dsfont}
\usepackage{xcolor}
\usepackage{xcolor}
\usepackage{color}
\usepackage{verbatim}
\usepackage{nomencl}
\usepackage[normalem]{ulem}
\usepackage{multirow}
\usepackage{makecell}
\usepackage{tikz}
\usetikzlibrary{arrows.meta, decorations.pathreplacing}
\DeclareUnicodeCharacter{2215}{/}
\DeclareUnicodeCharacter{0301}{\'{e}}
\usepackage[colorlinks=true,linkcolor=blue,citecolor=blue,urlcolor=blue]{hyperref}

\newcommand{\blue}[1]{\textcolor{blue}{#1}}

\begin{document}

\title{Nonequilibrium Dynamical Quantum Phase Transitions in the Extended Agassi Model}
\author{Ritesh Kumar Palai}
\email{riteshpalai297@gmail.com}
\affiliation{Department of Physics, \blue{Central University Of South Bihar}, Gaya, India}
\author{Rajiv Kumar}
\email{rajivkumar.rs.phy22@iitbhu.ac.in}
\affiliation{Department of Physics, \blue{Indian Institute of Technology (Banaras Hindu University)}, Varanasi - 221005, India}
\author{Venktesh Singh}
\email{venktesh@cusb.ac.in}
\affiliation{Department of Physics, \blue{Central University Of South Bihar}, Gaya, India}
\author{Sunil K. Mishra}
\email{sunilkm.app@iitbhu.ac.in}
\affiliation{Department of Physics, \blue{Indian Institute of Technology (Banaras Hindu University)}, Varanasi - 221005, India}

\begin{abstract}
We investigate dynamical quantum phase transitions (DQPTs) in the extended Agassi model following sudden quenches of  same level pairing ($\Sigma$),  monopole-monopole ($\chi$) and cross level pairing ($\Lambda$) strengths. Using exact diagonalization, we characterize the non-equilibrium dynamics through the Loschmidt echo and intensive return-rate function. Quenches across equilibrium phase boundaries produce cusp-like structures in the return rate that shows non-analyticities with increasing system size. Equilibrium finite size to large scaling identifies the $\chi$-driven transition as a second order phase transition similar to mean-field, Dicke, and Lipkin-Meshkov-Glick (LMG) models critical behaviour with the exponent of the order parameter $\beta \simeq 1/2$, while the $\Sigma$- and $\Lambda$-driven transitions are first order. We further characterize the associated dynamical singularities through a local scaling exponent $\alpha$, with $\alpha_\chi \simeq 2/3$, $\alpha_\Lambda \simeq 1/3$, and $\alpha_\Sigma \simeq 3/2$ of all three parameters. The distinct scaling behaviours across the parameter channels show that dynamical criticality is not determined solely by the equilibrium order of the transition, revealing multiple dynamical scaling regimes. These non-analyticities are associated with the Fisher zeros of the Loschmidt amplitude, which cross the real time axis at critical times and the time derivatives of the return-rate function to characterize the order of DQPTs within the extended Agassi model. 
\end{abstract}

\maketitle

\section{Introduction} 
Realistic nuclear many-body systems are governed by complex interactions, algebraic models have become increasingly valuable tools for gaining insight into their structure and dynamics over the past decade. These models have few degrees of freedom can often be solved exactly for large systems. These are valuable benchmarks for many-body behaviour in nuclear physics. They include the Elliott $\text{SU}(3)$ model~\cite{elliott1958collective} and the Interacting Boson model~\cite{iachello2006interacting}, which capture the rotational and collective degrees of freedom. Outside nuclear physics, analogous models such as the Jaynes-Cummings~\cite{Jaynes_1963}, Dicke~\cite{PhysRev.93.99} models of light-matter interaction, and the Lipkin-Meshkov-Glick (LMG) model~\cite{LIPKIN1965188} appear in quantum optics and condensed-matter systems. Evenly two-level models have proven useful, all of these models feature a collective dynamics and often exhibit dynamical quantum phase transitions (DQPT)~\cite{PhysRevLett.110.135704, heyl2018dynamical, Sachdev_1999, Sachdev_2011fcc}. The many-body systems driven out of equilibrium can exhibit phenomena that have no direct counterpart in their equilibrium behaviour. A central phenomenon of non-equilibrium dynamics is DQPT, in which time-dependent observables develop nonanalytic behaviour after quench~\cite{PhysRevLett.110.135704, PhysRevLett.115.140602, PhysRevB.109.224304, heyl2018dynamical, Sachdev_1999, Sachdev_2011fcc}. DQPT was introduced in the context of the Loschmidt amplitude and the associated rate function can develop non-analytically at a particular critical time, providing a dynamical analogue of the free energy at an equilibrium phase transition~\cite{PhysRevLett.110.135704}. The Fisher zeros of the Loschmidt amplitude in the complex-time plane, provide a natural framework for understanding DQPTs, this analogous to the partition function in equilibrium phase transitions \cite{PhysRevLett.110.135704, heyl2018dynamical, PhysRevB.107.094304, PhysRevB.110.224302, PhysRevResearch.5.033178}. In the thermodynamic limit, Fisher zeros may condense into continuous manifolds that approach and eventually pinch the real-time axis, confirming non-analyticities in the Loschmidt rate function at the critical times~\cite{PhysRevLett.110.135704, heyl2018dynamical, PhysRevResearch.5.033178}. However, for finite systems, the Fisher zeros remain discrete points, and the associated singularities are rounded. As the size of the system increases, the density of Fisher zeros also increases and their approach to the real-time axis becomes more pronounced, leading to progressively sharper signatures of dynamical criticality \cite{PhysRevLett.110.135704, heyl2018dynamical, PhysRevResearch.5.033178}. Since DQPTs have been investigated in quantum many-body spin systems \cite{PhysRevLett.115.140602, heyl2018dynamical, PhysRevB.109.224304}, fermionic models \cite{PhysRevB.105.094311, PhysRevB.103.224310}, long-range interacting systems, and systems with topology or symmetry-protected phases~\cite{PhysRevLett.115.140602, heyl2018dynamical, PhysRevB.89.125120, Zvyagin_4969869, PhysRevB.96.134427, PhysRevB.97.174401}.

The Agassi model \cite{AGASSI196849} is a two-level fermion system that combines a collective monopole-monopole strength with a Bardeen-Cooper-Schrieffer(BCS)-type pairing strength, extending the model by adding a pairing channel and based on an underlying algebra special orthogonal $\mathfrak{so}(5)$ \cite{iachello2006lie}. The pairing with quadrupole dynamics have rich physics with exactly solvable and allow us to explore symmetry-breaking in a controlled manner~\cite{DavisHeiss1986}. The Agassi model exhibits two broken-symmetry phases: one is associated with Hartree-Fock (parity-breaking) deformation, and the other is associated with the superconducting phase (pairing-breaking)~\cite{PhysRevC.97.054303, GarciaRamos2019, PhysRevC.95.064306}. Recently, the Agassi model has been extended by including an additional pairing term \cite{PhysRevC.97.054303, GarciaRamos2019, PhysRevC.106.064322}, refered as extended Agassi model, which couples pair correlated pairs across the two  levels, which is the extended Agassi model. This extended model has three independent tunable parameters and a richer phase structure, which contains both first-order and second-order phase transitions.

The non-equilibrium dynamics of the extended Agassi model have recently attracted attention to demonstrate that the correlation functions can distinguish different phases \cite{PhysRevC.106.064322}. This work motivates us to extend the analysis, such as DQPTs \cite{PEREZFERNANDEZ2022137133, PhysRevLett.110.135704, qc3w-s38g, PhysRevB.100.184313, PhysRevLett.118.015701, PhysRevB.89.125120, PhysRevB.104.115159, PhysRevB.96.224302, PhysRevB.109.224304}. Nevertheless, a systematic investigation of DQPTs is based on the Loschmidt amplitude and its associated rate function. DQPTs provide a framework for connecting non-equilibrium dynamics with equilibrium phases. For quenches across an equilibrium quantum phase, the evolution of the Loschmidt amplitude that reaches the real-time axis in the thermodynamic limit results in singularities in the dynamical rate function \cite{PhysRevLett.120.130601, PhysRevLett.110.135704, qc3w-s38g, PhysRevLett.115.140602, PhysRevB.107.094304, PhysRevB.110.224302,  PhysRevB.100.184313, PhysRevLett.118.015701, PhysRevB.89.125120, PhysRevB.104.115159,PhysRevB.109.224304}. The scaling properties associated with these singularities and established connections between dynamical critical behaviour and universality classes~\cite{PhysRevA.91.043409, qute_2018, heyl2018dynamical, PhysRevB.101.064427, PhysRevA.80.023810, PhysRevLett.93.237204}. 
Consequently, the study of DQPTs has become an important route for investigating critical phenomena in non-equilibrium many-body systems. These are relevant for schematic studies of collective phenomena in nuclear physics.

In this work, we investigate equilibrium and non-equilibrium analyses in the extended Agassi model, which includes monopole-monopole, same-level pairing, and cross-level pairing strengths, following quenches of its individual tunable parameters. We characterize the equilibrium phases and non-equilibrium dynamics using the Loschmidt amplitude and the associated return-rate function to identify dynamical critical times. By considering quenches across different parameter spaces, we examine dynamical singularities that depend on the underlying equilibrium phase transition, covering the same level pairing, monopole, and cross level pairing parameters. We further analyze the Fisher zeros of the Loschmidt amplitude, and investigate their evolution toward the real-time axis correlates with the emergence of nonanalytic structures in the return-rate function. We also analyze the finite-size scaling of the dynamical critical behaviour and extract the exponents with all parameter strengths. Our results reveal distinct scaling behaviour for the different quench parameters and compare between the equilibrium critical exponent and non-equilibrium dynamical exponents. Overall, these findings provide an equilibrium and dynamical perspective and extend the study from equilibrium to non-equilibrium quantum critical phenomena. This model offers a tractable algebraic platform for exploring dynamical criticality in the presence of competing strengths, while the distinct scaling behaviours to distinguish universal features of DQPTs in many-body systems.

The present study is organized as follows. In Sec.~\ref{Model}, we introduce the extended Agassi model and discuss its equilibrium phase diagram. In Sec.~\ref{scaling eqm}, we characterize the equilibrium phase transitions through finite-size scaling of the order parameters and extract the corresponding critical exponents, which are compared with mean-field predictions. In Sec.~\ref{losch_ech}, we investigate the nonequilibrium dynamics following sudden quenches of the $\Sigma$-, $\chi$-, and $\Lambda$-interaction parameters by analyzing the Loschmidt echo and the associated return-rate function, and identify the corresponding DQPTs. In Sec.~\ref{sd}, we extract the dynamical critical exponents from the singular behavior of the return-rate function and examine their scaling properties. In Sec.~\ref{FZ}, we study the Fisher-zero distributions and order of DQPTs of the Loschmidt amplitude in the complex-time plane and relate their evolution with system size to the emergence of dynamical non-analyticities.  Finally, Sec.~\ref{conclusion} summarizes the main results and outlines possible directions for future research.

%%%%%%%%%%%%%%%%%%%%%%%%%%%%%%%%%%%%%%%%%%%%%%%%%%%%%%%%%%%%%%%%%
\section{Modeling and Framework}
\label{Model}

\subsection{Extended Agassi Model }
The extended Agassi model describes $N$ fermions distributed in two energy levels, each level characterized by an even shell degeneracy $\Omega = 2j$, where $j$ is an integer associated with system size $N$. The total number of fermions is also related to $N_{\text{total}}$, i.e., $N_{\text{total}} = 2\Omega = 4j$. These two levels are separated by energy splitting $\epsilon$ and are indexed by the parities $\sigma = \pm 1$, ($\sigma = +1$, for the upper level, and $\sigma = -1$, for the lower level). The projection of angular momentum for each level is $m \in \{-j, -j+1, \dots, j\}$, excluding $m \neq 0$. Throughout this work, we restrict our analysis to half-filling, $N = 2j$, corresponding to $j$ fermion pairing. The Hamiltonian of the extended Agassi model \cite{PhysRevC.97.054303, GarciaRamos2019, PhysRevC.95.064306} in terms of fermion operators as:
\begin{equation} 
H = \epsilon J^0 - \mu \sum_{\substack{\sigma=\pm 1\\ \sigma'=\pm 1}}  A_\sigma^\dagger A_{\sigma'} -\frac{\xi}{2} \left[ (J^+)^2 + (J^-)^2 \right]
- 2\zeta A_0^\dagger A_0,
\label{eq.Ham}
\end{equation}
where \(\epsilon,\, \mu,\, \xi,\) and \(\zeta\) are tunable parameters associated with energy splitting, same-level pairing, monopole-monopole and cross-level pairing strengths, respectively. The fermion operators are generated by ten bilinear operators obeying the Lie algebra $\mathfrak{so}(5)$:
\begin{equation}
\begin{aligned}
J^{+} &= \sum_{m = -j}^{j} d_{1,m}^{\dagger} d_{-1,m} = (J^{-})^{\dagger}, \\ 
J^{0} &= \frac{1}{2}\sum_{m = -j}^{j} \left(d_{1,m}^{\dagger} d_{1,m} - d_{-1,m}^{\dagger} d_{-1,m}\right),  \\
A_{1}^{\dagger} &= \sum_{m = 1}^{j} d_{1,m}^{\dagger} d_{1,-m}^{\dagger} = (A_{1})^{\dagger}, \\ 
A_{-1}^{\dagger} &= \sum_{m = 1}^{j} d_{-1,m}^{\dagger} d_{-1,-m}^{\dagger} = (A_{-1})^{\dagger}, \\
A_{0}^{\dagger} &= \sum_{m = 1}^{j} \left(d_{-1,m}^{\dagger} d_{1,-m}^{\dagger} - d_{-1,-m}^{\dagger} d_{1,m}^{\dagger}\right) = (A_{0})^{\dagger}, \\ 
N_\sigma &= \sum_{m=-j}^{j} d_{\sigma,m}^\dagger d_{\sigma,m}, \qquad N = N_1 + N_{-1},
\label{eq.OP1}
\end{aligned}
\end{equation}
Here, the creation and annihilation operators $d_{\sigma,m}^{\dagger}$ ($d_{\sigma,m}$) satisfy the standard fermionic anti-commutation relations $\{d_{\sigma,m}, d_{\sigma',m'}^{\dagger}\} = \delta_{\sigma\sigma'} \delta_{mm'}$ and $\{d_{\sigma,m}, d_{\sigma',m'}\} = \{d_{\sigma,m}^{\dagger}, d_{\sigma',m'}^{\dagger}\}=0$. The underlying algebraic structure remains $\mathfrak{o}(5)$, which is generated by different operators, such as: six pair operators \(A\), three angular momentum operators \(J\), and the total particle number operator \(N\). The occupation level operators \(N_1\) and \(N_{-1}\) are not independent generators; they can be expressed as linear combinations of \(J_0\) and \(N\) \cite{GarciaRamos2019,AGASSI196849}. The Hamiltonian [\ref{eq.Ham}] is exact diagonalized in the seniority zero collective-pair basis (for a detailed description, see Appendix~\ref{app:hamiltonian}). We employ an orthogonal occupation number representation, with base states labeled \((n_-,n_+,n_0)\) satisfying \(n_-+n_++n_0=j\) \cite{PhysRevC.95.064306, GarciaRamos2019}. The operators of the Hamiltonian obey the following commutation relations as follows:

\begin{equation}
\begin{aligned}
\left[ J^+, J^- \right] &= 2 J^0, \quad \left[ J^0, J^\pm \right] = \pm J^\pm,\\
\left[ A_1, A_1^\dagger \right] &= j - N_1, \quad \left[ A_{-1}, A_{-1}^\dagger \right] = j - N_{-1},\\
\left[ A_0, A_0^\dagger \right] &= 2j - N, \quad \left[ A_0, A_1^\dagger \right] = -J^+,\\
\left[ A_0, A_{-1}^\dagger \right] &= -J^-, \quad \left[ A_1, A_0^\dagger \right] = -J^-,\\
\left[ A_{-1}, A_0^\dagger \right] &= -J^+, \quad \left[ J^+, A_1^\dagger \right] = 0,\\
\left[ J^+, A_0^\dagger \right] &= 2 A_1^\dagger, \quad \left[ J^+, A_{-1}^\dagger \right] = A_0^\dagger,\\
\left[ J^-, A_{-1}^\dagger \right] &= 0, \quad \left[ J^-, A_0^\dagger \right] = 2 A_{-1}^\dagger,\\
\left[ J^-, A_1^\dagger \right] &= A_0^\dagger, \quad \left[ J^0, A_1^\dagger \right] = A_1^\dagger,\\
\left[ J^0, A_{-1}^\dagger \right] &= -A_{-1}^\dagger, \quad \left[ J^0, A_0^\dagger \right] = 0.
\end{aligned}
\end{equation}
The total number of particles $N = N_1 + N_{-1}$ commutes with all ten generators, acting as a central element of Hamiltonian. Furthermore, the system has a discrete $\mathbb{Z}_2$ parity symmetry governed by the unitary parity operator $\Pi = e^{-i\pi J^0}$. Since $[H, \Pi] = 0$, all Hamiltonian eigenstates possess a well-defined parity eigenvalues $\pi = \pm 1$.

%%%%%%%%%%%%%%%%%%%%%%%%%%%%%%%%%%%%%%%%%%%%%%%%%%%%%%%%%%%%%
\subsection{Exploring Phase Transition}
In order to explore the different phase regimes in a phase diagram of a well-known Hamiltonian [Eq.~(\ref{eq.Ham})], these phases are associated with three tunable parameters. The Hamiltonian parameters are rescaled by the  factor $(2j-1)$ and introduce new dimensionless tunable parameters as \cite{PhysRevC.95.064306, GarciaRamos2019, DavisHeiss1986}:
\begin{equation}
\mu = \frac{\epsilon\Sigma}{2j - 1},\qquad
\xi = \frac{\epsilon\chi}{2j - 1},\qquad
\zeta = \frac{\epsilon\Lambda}{2j - 1}.
\label{eq:rescaled_parameters}
\end{equation}
These three parameters \(\Sigma\), \(\chi\), and \(\Lambda \ge 0\) control the strength of the same-level pairing, monopole-monopole, and cross-level pairing strengths, respectively\cite{PhysRevC.95.064306, GarciaRamos2019}. Expressed in terms of the rescaled parameters, the extended Agassi Hamiltonian [Eq.~(\ref{eq.Ham})] becomes
\begin{equation}
\begin{aligned}
H = &  \epsilon\Bigg[
J^{0} - \frac{\Sigma}{2j-1}\sum_{\sigma\sigma'} A_{\sigma}^{\dagger}A_{\sigma'} \\
& - \frac{\chi}{2(2j-1)}\bigl[(J^{+})^{2}+(J^{-})^{2}\bigr]
- \frac{2\Lambda}{2j-1} A_{0}^{\dagger}A_{0}
\Bigg].
\end{aligned}
\label{eq:H_scaled}
\end{equation}
This parametrization ensures that the tunable parameters of Hamiltonian [Eq.~(\ref{eq:H_scaled})] remain finite in the thermodynamic limit, rendering the mean-field and HFB energy surfaces well-defined. The phase diagram of the extended Agassi model presents five phases with different symmetry, parity, and additional pairing, as shown in Fig.~\ref{P1}. These phases with their boundaries are as follows\cite{PhysRevC.95.064306, GarciaRamos2019}:
\begin{figure}
    \centering
    \includegraphics[width=1.0\linewidth]{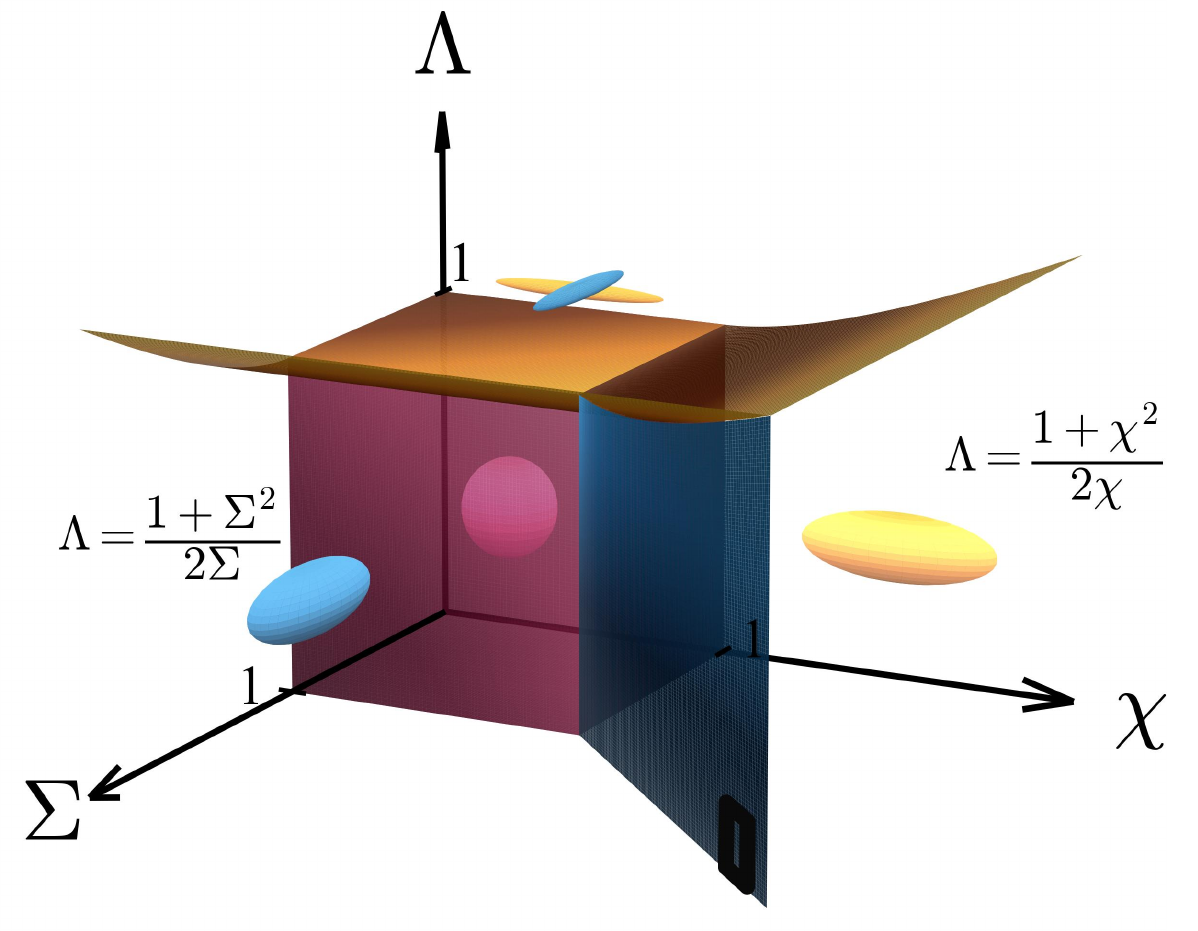}
   \caption{Phase diagram of the extended Agassi model. Second-order quantum phase transitions occur on the deep magenta vertical planes. First-order critical surfaces are indicated by the vibrant orange curved boundary, defined as $\Lambda=1$ for $\chi<1$ and $\Sigma<1$, $\Lambda=(1+\chi^2)/(2\chi)$ for $\chi>\Sigma$, and $\Lambda=(1+\Sigma^2)/(2\Sigma)$ for $\chi<\Sigma$, and by the deep cyan-blue vertical surface $\chi=\Sigma$ with $\Lambda<(1+\Sigma^2)/(2\Sigma)$. The symbols denote the symmetric (magenta sphere), HF deformed (orange oval), BCS deformed (blue oval), closed valley (black square outline), and combined HF-BCS deformed (crossed orange and cyan-blue ovals) phases Ref.\cite{PhysRevC.95.064306}.}
    \label{P1}
\end{figure}
 \begin{itemize}
   \item Symmetric or spherical solution:  \\        
  \( \text{\qquad \qquad} \chi < 1, \, \Sigma < 1, \, \Lambda < 1. \)\\
   \item HF deformed solution:  \\
  \( \text{\qquad \qquad} \chi > 1, \, \Sigma < \chi, \, \Lambda < \frac{1+\chi^2}{2\chi}. \)\\
  \item BCS deformed solution:  \\
  \( \text{\qquad \qquad} \chi < \Sigma, \, \Sigma > 1, \, \Lambda < \frac{1+\Sigma^2}{2\Sigma}. \)
   \item Combined HF-BCS deformed solution:  \\
   \( \text{\qquad \qquad} \Lambda > 1 \text{ for } \chi < 1, \, \Sigma < 1, \) \newline
   \( \text{\qquad \qquad} \Lambda > \frac{1+\chi^2}{2\chi} \text{ for } \chi > \Sigma > 1, \) \newline
   \( \text{\qquad \qquad} \Lambda > \frac{1+\Sigma^2}{2\Sigma} \text{ for } \Sigma > \chi > 1. \)
   \item Closed valley solution:  \\
  \( \text{\qquad \qquad} \chi = \Sigma, \, \Lambda < \frac{1+\Sigma^2}{2\Sigma}. \)
  \end{itemize}
Throughout this work, all analysis is performed for a system with pseudo-spin $j=300$, corresponding to $N=600$ fermions. Although the phase diagram is strictly valid in the thermodynamic limit $j\to\infty$, i.e. $j=300$, considered here, to capture the essential features of equilibrium and non-equilibrium dynamics for the exact diagonalization method of Hamiltonian. The choice $j=300$, is enough to exhibit collective behaviour and sharp DQPT features in the Loschmidt echo and return rate function. Still, the effects manifest as a rounding of cusp-like singularities, which become true nonanalytic only as $j\to\infty$. We also study the Loschmidt echo and return rate function for $j=8,100$ and $150$ mentioned in Appendix~\ref{AA2} to \ref{AA4}. However, key signatures, such as the linear scaling of the return rate near critical times are generate cusp like structure for $j=8$ mentioned in Appendix~\ref{AA2}, making it a robust minimal model to study the interplay between equilibrium criticality and non-equilibrium dynamics.

\section{Critical Exponents Across Equilibrium Regimes}
\label{scaling eqm}
\begin{figure}
    \centering
    \includegraphics[width=0.95\linewidth]{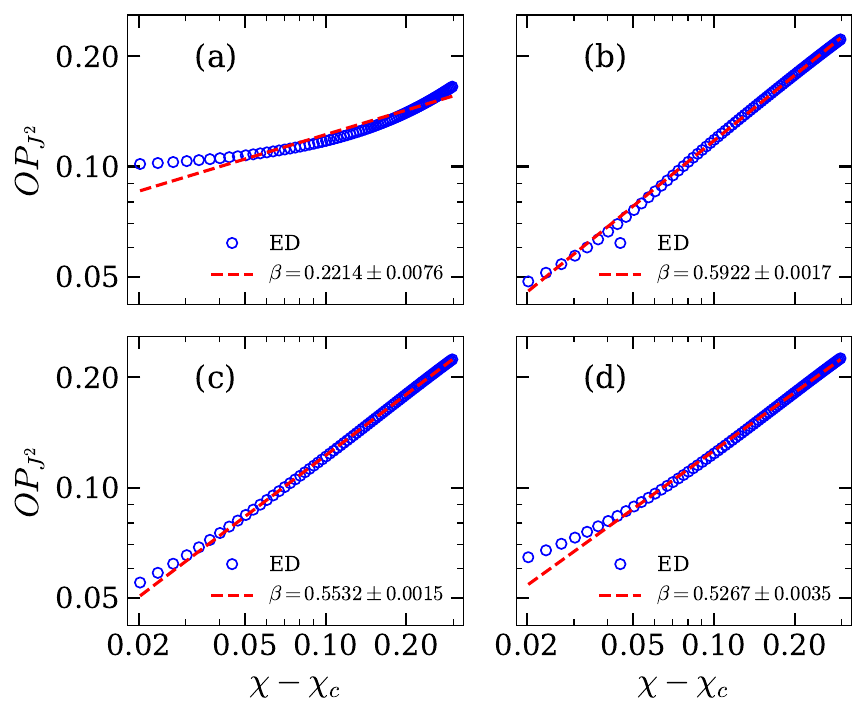}
    \caption{Scaling analysis of the equilibrium order parameter $\mathcal{OP}_{J^2}$ for the $\chi$-driven transition in the extended Agassi model (The $J^2$ refer to mathematical symmetry of coupling in mono-monopole term in Hamiltonian). Open circles denote exact-diagonalization (ED) results and dashed lines represent fits to $\mathcal{OP}_{J^2}\propto(\chi-\chi_c)^\beta$. Panels (a)--(d) correspond to $j=8$, $100$, $150$, and $300$, respectively. The critical couplings $\chi_c$ approach the thermodynamic critical value $\chi_c=1$, while the extracted critical exponents converge toward the mean-field value $\beta=1/2$. These results identify the symmetric-to-HF transition as a continuous quantum phase transition belonging to the mean-field universality class.}
    \label{Equilibrium Critical Exponent}
\end{figure}

In this section, we investigate the scaling properties of the extended Agassi model in the equilibrium profile. We perform a finite-size to large $j$ scaling analysis for different system sizes $j=8,100,150$ and $300$ to extract the critical exponents associated with the phase transitions for parameters $\chi, \Sigma$ and $\Lambda$ shown in Fig.~\ref{Equilibrium Critical Exponent}. This analysis of static critical behaviour provides insight into the universality properties \cite{PhysRevLett.115.140602}. Although equilibrium phase transitions exhibit scaling behavior consistent with the LMG and Dicke universality classes \cite{J.Vidal_2006, PhysRevA.80.023810, PhysRevLett.93.237204},the extended Agassi model reveals critical features that are both common to the LMG and Dicke universality classes.
The phase diagram of the extended Agassi model [Eq.~(\ref{eq:H_scaled})] exhibits five distinct phases. These phases are separated by critical surfaces, some of which are continuous (second-order) transitions, while the remaining phases are discontinuous (first-order) transitions. To characterize these phase transitions via a finite-size scaling analysis of the order parameter $\mathcal{OP}_{J^2}$.

%%%%%%%%%%%%%%%%%%%%%%%%%%%%%%%%%%%%%%%%%%%%%%
\subsection{Equilibrium Critical Exponents order parameter}
\label{sec:eq_crit_exp_op}
The equilibrium critical exponent $\beta$ is extracted from the scaling relation of the order parameters $\mathcal{OP}_{J^2}$ as

\begin{equation}
\mathcal{OP}_{J^2}\propto |g-g_c|^{\beta},
\end{equation}
where $g\in\{\chi,\Sigma,\Lambda\}$ denotes the tunable parameters and $g_c$ are the corresponding critical points. Fig.~\ref{Equilibrium Critical Exponent} shows the scaling behaviour of $\mathcal{OP}_{J^2}$ for the $\chi$-driven transition. As the system size increases from $j=8$ to $j=300$, the critical exponent evolves in a particular range, from $\beta=0.2214\pm0.0076$ to $\beta=0.5267\pm0.0035$, approaching the mean-field value $\beta=1/2$. Simultaneously, the finite-size critical coupling $\chi_c$ converges to the thermodynamic value $\chi_c=1$. The exponents obtained for $j=100$, $150$, and $300$ are already close to the mean-field prediction, demonstrating systematic convergence toward the universality class of the LMG and Dicke models \cite{PhysRevB.28.3955, PhysRevE.91.062108, PhysRevE.70.066130, PhysRevA.80.023810, PhysRevLett.93.237204, Bapst_2012}.

\begin{table}
\centering
\caption{Finite-size and large $j$ scaling of the equilibrium order-parameter critical exponent $\beta$ for the $\chi$-, $\Sigma$-, and $\Lambda$-driven transitions in the extended Agassi model. The exponent converges to the mean-field value $\beta = 1/2$ \cite{PhysRevE.91.062108,PhysRevE.70.066130,PhysRevA.80.023810,PhysRevLett.93.237204,Bapst_2012}for the continuous $\chi$-driven transition, while the vanishing exponent $\beta \approx 0$ confirms the first-order nature of the $\Sigma$- and $\Lambda$-driven transitions.}
\label{tab:equilibrium_beta}
\setlength{\tabcolsep}{3pt}        
\renewcommand{\arraystretch}{1.8} 
\resizebox{1.0\columnwidth}{!}{%
\begin{tabular}{c|cc|cc|cc}
\hline\hline
$j$ & \multicolumn{2}{c|}{$\chi$}
    & \multicolumn{2}{c|}{$\Sigma$}
    & \multicolumn{2}{c}{$\Lambda$} \\
\cline{2-7}
    & $\chi_c$ & $\beta$
    & $\Sigma_c$ & $\beta$
    & $\Lambda_c$ & $\beta$ \\
\hline
8   & 1.45000 & $0.2214\pm0.0076$
    & 1.00000 & $0.0018\pm0.0001$
    & --      & -- \\
100 & 1.10333 & $0.5922\pm0.0017$
    & 1.06500 & $0.0001\pm0.0000$
    & 1.73000 & $0.0003\pm0.0000$ \\
150 & 1.08000 & $0.5532\pm0.0015$
    & 1.42500 & $0.0001\pm0.0000$
    & 1.02000 & $0.0007\pm0.0000$ \\
300 & 1.05000 & $0.5267\pm0.0035$
    & 1.55500 & $0.0000\pm0.0000$
    & 1.09500 & $0.0004\pm0.0000$ \\
\hline\hline
\end{tabular}
}
\end{table}
In contrast, the phase transitions depending on $\Sigma$ and $\Lambda$ exhibit vanishing critical exponents for all system sizes considered. The finite size scaling of the order parameter remains always near zero, indicating the absence of a continuous power-law onset of the phases. Instead, the order parameter changes discontinuously at the transition point  \cite{GarciaRamos2019}, which is the hallmark of a first-order quantum phase transition. The small deviations from zero originate from finite-size rounding and decrease as the system size increases. The resulting critical exponents for all parameters are summarized in Table~\ref{tab:equilibrium_beta}. These results establish two distinct phase classes of critical equilibrium behaviour in the extended Agassi model. The $\chi$ depending symmetric-to-HF transition is continuous and belongs to the mean-field, LMG, Dicke universality class with $\beta=1/2$, while the $\Sigma$ and $\Lambda$ depending transitions correspond to first-order quantum phase transitions characterized by $\beta\simeq0$. The coexistence of both continuous and discontinuous transitions within the same exactly solvable many-body model provides an ideal setting for comparing equilibrium and non-equilibrium critical phenomena.
%%%%%%%%%%%%%%%%%%%%%%%%%%%%%%%%%%%%%%%%%%%
\subsection{Comparison with the Dicke and Lipkin Meshkov Glick Model}
The continuous $\chi$-driven transition in the extended Agassi model shares the same universality class as the superradiant transition in the Dicke model and the ferromagnetic transition in the LMG model. For the Dicke model, the order parameter exponent is obtained to be $\beta = 1/2$, which is the same as that in the LMG model, suggesting the same universality \cite{PhysRevA.80.023810, PhysRevLett.93.237204}. Furthermore, it has been shown that the second-order phase transitions of the LMG  belong to the same universality class, and arguably those of the Dicke model \cite{PhysRevA.91.043409}. These models share the same underlying algebraic structure: they are described by collective operators and satisfy the $\mathfrak{so}(3)$ and $\mathfrak{so}(5)$ Lie algebra~\cite{iachello2006lie}.
While the equilibrium critical behaviour is well understood within the framework of mean-field universality, the non-equilibrium dynamics following sudden parameter quenches, which is generally governed by a distinct universality class is not explored. To address this, we next explored the DQPTs and the scaling properties to extract the corresponding dynamical critical exponents.

\begin{figure}
    \centering
    \includegraphics[width=1.0\linewidth, height=0.85\linewidth]{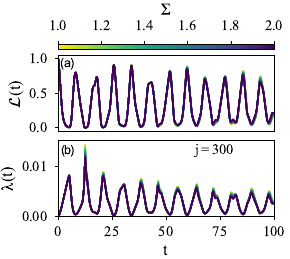}
  \caption{Loschmidt dynamics in the extended Agassi model following a sudden $\Sigma$ quench for $j=300$. (a) Loschmidt echo $\mathcal{L}(t)$ and (b) return-rate function $\lambda(t)$ for a quench from $H_i(\Sigma_i,\chi_i,\Lambda_i) = H(0.5,0.5,0.0)$  to  ($\Sigma_f \in [1.0, 2.0]$, $\chi_f=1.5$, $\Lambda_f=0.5$). The quench crosses the equilibrium HF--BCS phase boundary at $\Sigma_c^{\rm eq}=1.5$. A pronounced signature of a DQPT appears at $\Sigma_f=1.0$.}
    \label{sig_los}
\end{figure}

%%%%%%%%%%%%%%%%%%%%%%%%%%%%%%%%%%%%%%%%%%%%%%%%%%%%%%%%%%%%
\section{Non-equilibrium Loschmidt Dynamics and DQPT's}
\label{losch_ech}
The non-equilibrium dynamics of the extended Agassi model following sudden quenches of the control parameters $\Sigma, \chi$, and $\Lambda$ can be characterized through the Loschmidt echo \cite{PhysRevLett.110.135704, qc3w-s38g, PhysRevB.100.184313, PhysRevLett.118.015701, PhysRevB.89.125120, PhysRevB.104.115159, PhysRevB.109.224304} as
\begin{equation}
\mathcal{L}(t) = |g(t)|^2 = \left| \langle \psi_i|e^{-iHt}|\psi_i\rangle \right|^2,
\end{equation}
where $|\psi_i\rangle$ denotes the initial ground state of the pre-quench Hamiltonian $H_i$ and $H_f =H$ is the post-quench Hamiltonian. To identify DQPTs, we analyze the intensive return-rate function \cite{PhysRevLett.120.130601, PhysRevLett.115.140602, PhysRevLett.110.135704, qc3w-s38g, PhysRevB.100.184313, PhysRevLett.118.015701, PhysRevB.89.125120, PhysRevB.104.115159, PhysRevB.109.224304} as
\begin{equation}
\lambda(t) = -\lim_{N\to\infty} \frac{1}{N} \ln \mathcal{L}(t),
\label{LR1}
\end{equation}
whose non-analytic behaviour for large $j$, signals the occurrence of a DQPTs. We now analyze the DQPTs for all tunable parameters $\chi, \Sigma$ and $\Lambda$ of the Hamiltonian.

%%%%%%%%%%%%%%%%%%%%%%%%%%%%%%%%%%%%%%%%%%%%%%%%%%%%%
\subsection{Same-level Pairing (\texorpdfstring{$\Sigma$}{Sigma})-driven DQPT}
The Loschmidt dynamics of the extended Agassi model for the system size $j=300$ shown in Fig.~\ref{sig_los}. The system is initially prepared in the symmetric phase at $H_i=H(\Sigma_i,\chi_i,\Lambda_i)=(0.5,0.5,0.0)$ and suddenly quenched to a final Hamiltonian $H_f$ with fixed $\chi_f=1.5$ and $\Lambda_f=0.5$, while the same-level pairing strength $\Sigma_f$ is varied within the range $1.0\leq\Sigma_f\leq2.0$. This quench crosses the first-order equilibrium HF transition point to the BCS located at $\Sigma_c^{\rm eq}=1.5$. The Loschmidt echo $\mathcal{L}(t)$ exhibits a sequence of coalescence and revival oscillations, while the corresponding return-rate function $\lambda(t)$ develops increasingly sharp cusp-like structures. For $j=300$, the growing density of collective many-body states contributes to the post-quench dynamics effects, producing a pronounced signature of a DQPT. The strongest cusp in $\lambda(t)$ occurs for $\Sigma_f=1.0$, well separated from the equilibrium critical value $\Sigma_c^{\rm eq}=1.5$. This demonstrates that dynamical criticality is not determined by the equilibrium phase diagram but is governed by the overlap of the initial state with the spectrum of the post-quench Hamiltonian. 

\begin{figure}
    \centering
    \includegraphics[width=1.0\linewidth, height=0.85\linewidth]{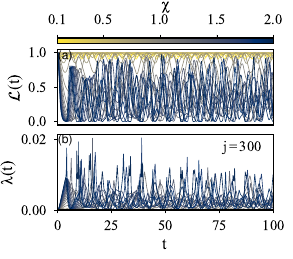}
    \caption{Loschmidt dynamics in the extended Agassi model following a sudden $\chi$ quench for $j=300$. (a) Loschmidt echo $\mathcal{L}(t)$ and (b) return-rate function $\lambda(t)$ for an initial state prepared at $H_i=H(\Sigma_i,\chi_i,\Lambda_i)=(0.5,0.5,0.0)$ and evolved under a final Hamiltonian with $\Sigma_f=0.5$ and $\Lambda_f=0$, while varying $\chi_f\in[0.5,2.5]$. The quench trajectory crosses the equilibrium symmetric--HF phase boundary at $\chi_c^{\rm eq}=1$. The strongest signature of DQPT occurs at $\chi_f=1.367$.}
    \label{R5}
\end{figure}

\subsection{Monopole-Monopole(\texorpdfstring{$\chi$}{chi}) driven DQPT}
The Loschmidt echo $\mathcal{L}(t)$ for $j=300$ following sudden quenches in the monopole-monopole strength $\chi$  shown in Fig.~\ref{R5}. The system is initially prepared in the symmetric phase at $H_i(\Sigma_i,\chi_i,\Lambda_i) = H(0.5,0.5,0.0)$ and evolved under a final Hamiltonian with fixed parameters $\Sigma_f=0.5$ and $\Lambda_f=0$, while $\chi_f$ is varied within the range $0.5\leq\chi_f\leq2.5$. This quench protocol crosses the equilibrium symmetric-HF transition point located at $\chi_c^{\rm eq}=1$. The $\mathcal{L}(t)$ exhibits a sequence of coalescence and revival oscillations whose structure becomes strongly dependent on the final strength of $\chi$. Correspondingly, the $\lambda(t)$ develops sharp  peaks arising from many collective eigen states participating in the post-quench dynamics. The strongest DQPT precursor is observed at $\chi_f=1.397$, where the $\lambda(t)$ displays its most pronounced singular structure. In particular, this dynamical critical value is well within the HF phase and is displaced from the equilibrium critical coupling $\chi_c^{\rm eq}=1$. This separation of the dynamical critical point is governed by the overlap of the initial state with the excitation spectrum of the post-quench Hamiltonian rather than by equilibrium criticality.

\begin{figure}
    \centering
    \includegraphics[width=1.0\linewidth, height=0.85\linewidth]{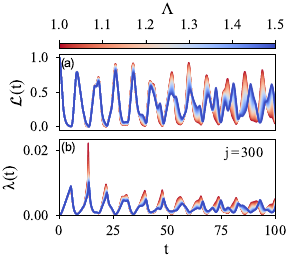}
    \caption{Loschmidt dynamics in the extended Agassi model following a sudden $\Lambda$ quench for $j=300$. (a) Loschmidt echo $\mathcal{L}(t)$ and (b) return-rate function $\lambda(t)$ for an initial state prepared at $H_i(\Sigma_i,\chi_i,\Lambda_i)=H(0.5,0.5,0.0)$ and evolved under a final Hamiltonian with $(\Sigma_f,\chi_f)=(2.0,1.5)$ while varying $\Lambda_f\in[1.0,1.5]$. The quench trajectory crosses the equilibrium first-order BCS--HF--BCS phase boundary at $\Lambda_c^{\rm eq}=1.25$. The strongest signature of DQPT occurs at $\Lambda_f = 1.0$.}
    \label{R3}
\end{figure}

\subsection{Cross--level(\texorpdfstring{$\Lambda$}{Lambda}) pairing-driven DQPT}
The Loschmidt dynamics for $j=300$ after sudden quenches in cross-level pairing strength $\Lambda$ shown in Fig.~\ref{R3}. The system is initially prepared in the symmetric phase at $H_i$ and evolved under a final Hamiltonian with parameters $(\Sigma_f,\chi_f)=(2.0,1.5)$, while $\Lambda_f$ is varied within the range $1.0\leq\Lambda_f\leq1.5$. This quench trajectory crosses the first-order equilibrium BCS-HF-BCS transition point located at $\Lambda_c^{\rm eq}=1.25$. From Fig.~\ref{R3}, $\mathcal{L}(t)$ exhibits pronounced coalescence-and-revival oscillations, while the corresponding return-rate function $\lambda(t)$ develops increasingly sharp cusp-like structures. For $j=300$, these features constitute the signature of DQPT. The singular behaviour is observed for values of $\Lambda_f$, indicating that the location of the dynamical critical point is governed by the post-quench excitation spectrum and the overlap distribution of the initial state rather than by the equilibrium phase point. The persistence of coherent revivals with time reflects the discrete collective spectrum associated with the  dynamical behaviour of the extended Agassi model. These results demonstrate that quenches in the cross-level pairing channel provide an independent route to non-equilibrium criticality, complementing the behaviour observed for HF and same-level pairing quenches.

The results presented above for all three tunable parameters demonstrate that the extended Agassi model exhibits non-analytic signatures of DQPTs under quenches across individual parameter, namely the same level pairing $\Sigma$, the monopole-monopole $\chi$ and the cross level pairing $\Lambda$. In each case, the Loschmidt echo and  return-rate function develops pronounced cusp-like structures, providing a signature of nonanalytic behaviour in the thermodynamic limit. A common feature of all three quench protocols is that the dynamical signatures occur at parameter values that are shifted from the corresponding equilibrium transition points. This observation indicates that the non-equilibrium critical behaviour is governed by the interplay between the initial-state preparation and the excitation spectrum of the post-quench Hamiltonian rather than by equilibrium criticality. Moreover, the distinct temporal structures observed for the three parameter channels suggest that the associated dynamical behaviour may belong to different scaling regimes. These findings naturally motivate a quantitative characterization of both equilibrium and non-equilibrium criticality through the extraction of critical exponents, which is the focus of the next section.

%%%%%%%%%%%%%%%%%%%%%%%%%%%%%%%%%%%%%%%%%%%%%%%%%%%%%%%%%%%%%%%%%
\section{Critical Exponents Across Non-equilibrium Regimes}
\label{sd}
\begin{figure}
    \centering
    \includegraphics[width=1.0\linewidth, height=0.9\linewidth]{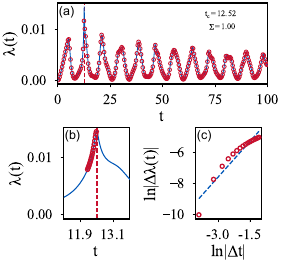}
   \caption{The dynamical critical exponent for the $\Sigma$-driven quench at $j=300$. (a) Return-rate function $\lambda(t)$. (b) Enlarged view of the strongest DQPT precursor near the critical time $t_c=12.52$. (c) Log--log scaling of $|\lambda(t)-\lambda(t_c)|$ versus $|t-t_c|$, demonstrating the power-law behavior used to extract the dynamical exponent $\alpha_\Sigma$. Open circles denote numerical data and the dashed line represents the linear fit. The quench is performed from $H_i(\Sigma_i,\chi_i,\Lambda_i) = H(0.5,0.5,0.0)$ to $H_f(\Sigma_f,\chi_f,\Lambda_f) = (1.0,1.5,0.5)$. }
    \label{R2}
\end{figure}

\begin{figure}
    \centering
    \includegraphics[width=1.0\linewidth, height=0.9\linewidth]{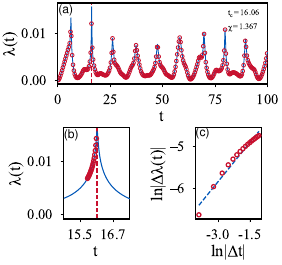}
    \caption{The dynamical critical exponent for the $\chi$-driven quench at $j=300$. (a) Return-rate function $\lambda(t)$. (b) Enlarged view of the strongest DQPT signature near the critical time $t_c=16.06$. (c) Log--log scaling of $|\lambda(t)-\lambda(t_c)|$ versus $|t-t_c|$, demonstrating the power-law behavior used to extract the dynamical exponent $\alpha_\chi$. Open circles denote numerical data and the dashed line represents the linear fit.The quench is performed from \(H_i(\Sigma_i,\chi_i,\Lambda_i)=H(0.5,0.5,0.0)\)
   to \(H_f(\Sigma_f,\Lambda_f)=(0.5,0.0)\) and \(\chi_f=1.367\).}
    \label{R6}
\end{figure}

In this section, we investigate scaling properties of the return rate function to obtain critical exponents  in the extended Agassi model in a non-equilibrium profile. We perform a scaling analysis from finite-size $j$ to large $j$, i.e., $j=8,100,150$ and $300$ to extract the critical exponents associated with the return-rate function $\lambda(t)$ shown in Figs.~\ref{R2} to ~\ref{R4}. We focus on the intensive return rate function $\lambda(t)$ [Eq.~(\ref{LR1})], which in the thermodynamic limit satisfies the large-deviation form $\mathcal{L}(t)\sim e^{-j\lambda(t)}$, which identifies $\lambda(t)$ as the dynamical analogue of a free-energy \cite{PhysRevB.98.184303}. Non-analytical in $\lambda(t)$ at critical times $t_c$, provide direct signatures of DQPTs. This exponents enables a comparison between static and dynamical behaviour and provides insight into the universality properties of the model \cite{PhysRevLett.115.140602, PhysRevA.91.043409}. However, the non-equilibrium transitions exhibit distinct dynamical exponents that depend on the tunable parameters $\Sigma, \chi$ and $\Lambda$ driving the quench. The resulting analysis reveals contrasting features of criticality in the non-equilibrium dynamics.

To quantify the local structure of these dynamical singularities, we introduce a dynamical critical exponent $\alpha_p$ \cite{PhysRevB.104.115159, PhysRevB.107.L100419} through the scaling relation
\begin{equation}
|\lambda(t)-\lambda(t_c)| \propto |t-t_c|^{\alpha_p},
\label{eq:dynamical_scaling}
\end{equation}
where $t_c$ denotes the critical time associated with the DQPTs signature. The exponent $\alpha_p$, (where $p=\Sigma,\chi,\Lambda$) characterizes the non-analytic structures and provides a quantitative measure of critical behaviour. Numerically, $\alpha_p$ is extracted from a linear fit of
\begin{equation}
\ln |\lambda(t)-\lambda(t_c)| = \alpha_p \ln |t-t_c| + C,
\end{equation}
carried out within a suitable scaling law around $t_c$ and $C$ is scaling constant. By comparing the resulting exponents for $\Sigma$, $\chi$, and $\Lambda$ driven quenches, we examine whether the non-equilibrium critical behaviour follows the equilibrium mean-field universality class or exhibits distinct dynamical scaling properties~\cite{PhysRevA.91.043409}.

\subsection{\texorpdfstring{$\Sigma$}{Sigma}-Dynamical Critical Exponent}
The dynamical critical behaviour associated with quenches of the same level pairing strength $\Sigma$ is shown in Fig.~\ref{R2} for system size $j=300$. Panel (a) presents the return-rate function $\lambda(t)$, while panel (b) displays a zoomed view of the dominant cusp-like structure at the critical time $t_c=12.52$. The corresponding log--log fitting of panel (b) shown in panel (c) reveals the signature of the power-law behaviour of $|\lambda(t)-\lambda(t_c)|$ in $|t-t_c|$ in the vicinity of the singularity.

Fitting the scaling relation [Eq.~(\ref{eq:dynamical_scaling})] for $\alpha_p = \alpha_\Sigma$, yields the dynamical exponent $\alpha_\Sigma = 1.5373 \pm 0.1243.$ This exponent is numerically nearby $3/2$. The finite-size analysis of $\alpha_\Sigma$, together with the associated critical time and quench parameters for system size  $j=8$, $100$, and $150$ mentioned in the Appendix~\ref{AA2}. As $j$ increases, the cusp in $\lambda(t)$ becomes increasingly sharp, indicating the gradual emergence of a non-analytic dynamical singularity in the thermodynamic limit. The $\Sigma$-driven DQPT originates from quenches of the same-level pairing strength that generates BCS correlations in the post-quench state. Unlike the continuous symmetric-HF transition associated with the $\chi$ equilibrium transition, it is of the first order. The different value of $\alpha_\Sigma$ from equilibrium therefore suggests that the local structure of the dynamical singularity depends sensitively on the parameter channel driving the quench and may retain signatures of the equilibrium transition. Although the numerical value $\alpha_\Sigma \simeq 3/2$ suggests a distinct asymptotic scaling behaviour, the identification of a class of dynamical universality requires a systematic extrapolation of finite-size and analysis closer to the thermodynamic limit. The overall analysis for all $j$, are summarized in Table~\ref{tab:dqpt_exponents}.

\begin{table*}
\centering
\caption{Critical quantities associated with the strongest DQPTs precursors in the extended Agassi model for system sizes $j=8$, $100$, $150$, and $300$. The system is initialized in the symmetric phase at $H_i$$(\Sigma_i,\chi_i,\Lambda_i)=(0.5,0.5,0.0)$. For the $\Sigma$-driven quenches, the final parameters are $H_f$ $(\Sigma_f,\chi_f,\Lambda_f)=(1.0,1.5,0.5)$, $(1.655,1.5,0.5)$, $(1.103,1.5,0.5)$, and $(1.0,1.5,0.5)$ for $j=8$, $100$, $150$, and $300$, respectively. For the $\Lambda$-driven quenches, the final parameters are $(2.0,1.5,1.172)$, $(2.0,1.5,1.241)$, $(2.0,1.5,1.259)$, and $(2.0,1.5,1.0)$, while for the $\chi$-driven quenches the final couplings are $\chi_f=1.603$, $2.041$, $1.259$, and $1.367$, respectively, with $\Sigma_f=0.5$ and $\Lambda_f=0$. The quantities $t_c$, $\lambda(t_c)$, and $\alpha$ denote the critical time, the return-rate function evaluated at the critical time, and the dynamical singularity exponent extracted from the scaling relation $|\lambda(t)-\lambda(t_c)|\propto|t-t_c|^{\alpha}$.}

\label{tab:dqpt_exponents}
\begin{tabular}{c|cc|cc|cc}
\hline\hline
$j$
& \multicolumn{2}{c|}{$\Sigma$(First order)}
& \multicolumn{2}{c|}{$\Lambda$(First order)}
& \multicolumn{2}{c}{$\chi$(Second order)} \\
\cline{2-7}
& $t_c$ & $\alpha$
& $t_c$ & $\alpha$
& $t_c$ & $\alpha$ \\
\hline
8
& 26.099220 & $1.613273\pm0.028122$
& 16.053211 & $0.382480\pm0.015640$
& 20.554111 & $0.535747\pm0.035817$
\\
100
& 90.938188 & $0.353894\pm0.013050$
& 59.691938 & $0.129949\pm0.001373$
& 30.42608522 & $0.599597\pm0.031462$
\\
150
& 20.904181 & $0.204428\pm0.004527$
& 64.452891 & $0.591008\pm0.057832$
& 6.161232 & $0.646551\pm0.036961$
\\
300
& 12.522504 & $1.537306\pm0.124286$
& 13.262652 & $0.319059\pm0.012378$
& 16.06321264 & $0.622468\pm0.028502$
\\
\hline\hline
\end{tabular}
\end{table*}

\subsection{\texorpdfstring{$\chi$}{chi}- Dynamical Critical Exponent}
The dynamical critical behaviour associated with quenches of $\chi$ is illustrated in Fig.~\ref{R6} for system size, $j=300$. Panel (a) shows the return-rate function $\lambda(t)$, while panel (b) displays a zoomed view of the cusp-like structure at the critical time $t_c=16.06$. The corresponding log-log fitting of panel (b) presented in panel (c) demonstrates a power-law behaviour of $|\lambda(t)-\lambda(t_c)|$ in $|t-t_c|$ near the singularity.

Fitting the scaling relation [\ref{eq:dynamical_scaling}] for $\alpha_p = \alpha_\chi$ yields the dynamical critical exponent $\alpha_\chi = 0.6225 \pm 0.0285$. The exponent value lies nearly $2/3$, although we interpret it as an asymptotic exponent rather than evidence of a distinct universality class. The finite-size evolution of $\alpha_\chi$, together with the corresponding critical times and quench parameters for $j=8,100$ and $150$ mentioned in the Appendix~\ref{AA3}. As the size of the system increases from $j=8$ to $300$, the cusp in $\lambda(t)$ becomes progressively sharper, indicating the approach toward a non-analytic approach in the thermodynamic limit. Unlike the equilibrium $\chi$-driven transition, which is characterized by the mean-field order-parameter exponent $\beta=1/2$. DQPTs manifest as non-analytic in the rate function, where the Loschmidt echo develops exact zeros at critical times. The overall analysis for all $j$, are summarized in Table~\ref{tab:dqpt_exponents}.

\begin{figure}
    \centering
    \includegraphics[width=1.0\linewidth, height=0.9\linewidth]{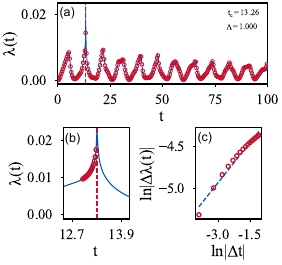}
   \caption{The dynamical critical exponent for the $\Lambda$-driven quench at $j=300$. (a) Return-rate function $\lambda(t)$. (b) Enlarged view of the strongest DQPT precursor near the critical time $t_c=13.26$. (c) Log--log scaling of $|\lambda(t)-\lambda(t_c)|$ versus $|t-t_c|$, demonstrating the power-law behavior used to extract the dynamical exponent $\alpha_\Lambda$. Open circles denote numerical data and the dashed line represents the linear fit. The quench is performed from
   \(H_i(\Sigma_i,\chi_i,\Lambda_i)=H(0.5,0.5,0.0)\)
   to \(H_f(\Sigma_f,\chi_f)=(0.5,0.5)\) and \(\Lambda_f=1.0\).}
    \label{R4}
\end{figure}

\subsection{\texorpdfstring{$\Lambda$}{Lambda}-Dynamical Critical Exponent}
The dynamical critical behaviour associated with quenches in the cross level pairing strength $\Lambda$ is shown in Fig.~\ref{R4} for system size $j=300$. Panel (a) displays the return-rate function $\lambda(t)$, while panel (b) presents a zoomed view of the dominant cusp-like structure at the critical time $t_c=13.26$. The corresponding log-log scaling of panel (b) shown in panel (c) reveals a clear power-law behaviour of $|\lambda(t)-\lambda(t_c)|$ on $|t-t_c|$ near the singularity.

Using the scaling relation [Eq.~(\ref{eq:dynamical_scaling})] for $\alpha_p = \alpha_\Lambda $ yields the dynamical critical exponent $\alpha_\Lambda = 0.3191 \pm 0.0124$. This exponent value lies nearly $1/3$ and differs significantly from both $\chi$-driven, i.e., $\alpha_\chi=0.6225\pm0.0285$ and $\Sigma$-driven, i.e., $\alpha_\Sigma=1.5373\pm0.1243$. The value $\alpha_\Lambda \simeq 1/3$ represents a distinct asymptotic exponent associated with the cross level pairing strength. The finite-size evolution of $\alpha_\Lambda$, together with the associated critical times and quench parameters for $j=8$, $100$, and $150$ are mentioned in Appendix~\ref{AA4}. As the size of the system increases from $j=8$ to $300$ , the cusp in the return-rate function becomes progressively sharper, indicating the approach toward a genuine non-analytic in the thermodynamic limit. The $\Lambda$ strength couples to the correlated-pair sector of the underlying $\mathfrak{so}(5)$ algebra and therefore probes a dynamical channel distinct from both the monopole ($\chi$) and the same level pairing ($\Sigma$) strengths. Remarkably, although both the $\Sigma$- and $\Lambda$-driven equilibrium transitions are first order, their dynamical critical exponents differ substantially. This observation suggests that the local structure of the Loschmidt singularity is determined not solely by the equilibrium order of the transition but also by the specific excitation and overlap structure generated by the quench. The overall analysis for all $j$ with individually all three parameters are summarized and comparison is given in Table~\ref{tab:dqpt_exponents}. To further characterize the nature of the dynamical singularities, we complement this exponent analysis with an examination of the Fisher zeros and the derivative structure of the rate function.

\begin{figure*}
    \centering
    \includegraphics[width=0.3\linewidth, height=0.28\linewidth]{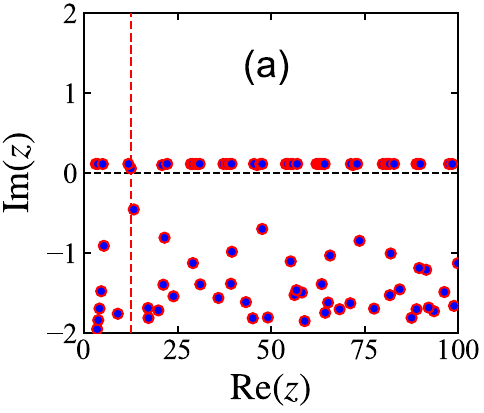}
    \includegraphics[width=0.3\linewidth, height=0.28\linewidth]{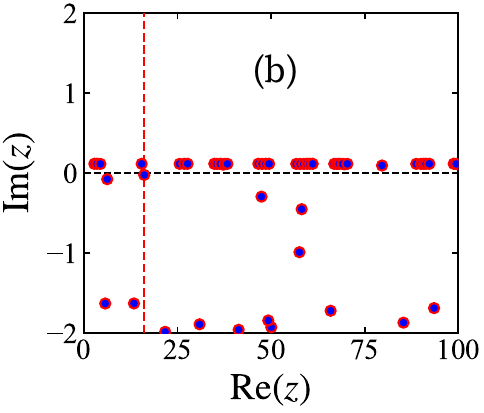}
    \includegraphics[width=0.3\linewidth, height=0.28\linewidth]{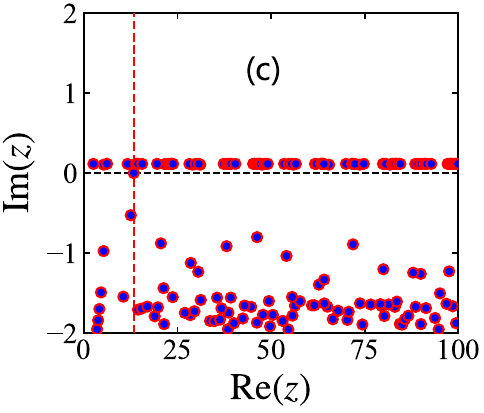}
  \caption{Fisher zeros of the Loschmidt amplitude in the complex-time plane for \(j=300\) in the  following quenches of parameters (a) \(\Sigma_f=1.0\), (b) \(\chi_f=1.367\), and (c) \(\Lambda_f=1.0\). The horizontal dashed line denotes the real-time axis, \(\operatorname{Im}z=0\), and the vertical dashed lines indicate the corresponding dynamical critical times \(t_c\simeq12.52\), \(16.06\), and \(13.26\), respectively. The critical times are identified from the real-time dynamical analysis and are compared with the corresponding Fisher-zero distributions.}
    \label{FS}
\end{figure*}

\begin{figure*}
    \centering
    \includegraphics[width=0.9\linewidth, height=0.5\linewidth]{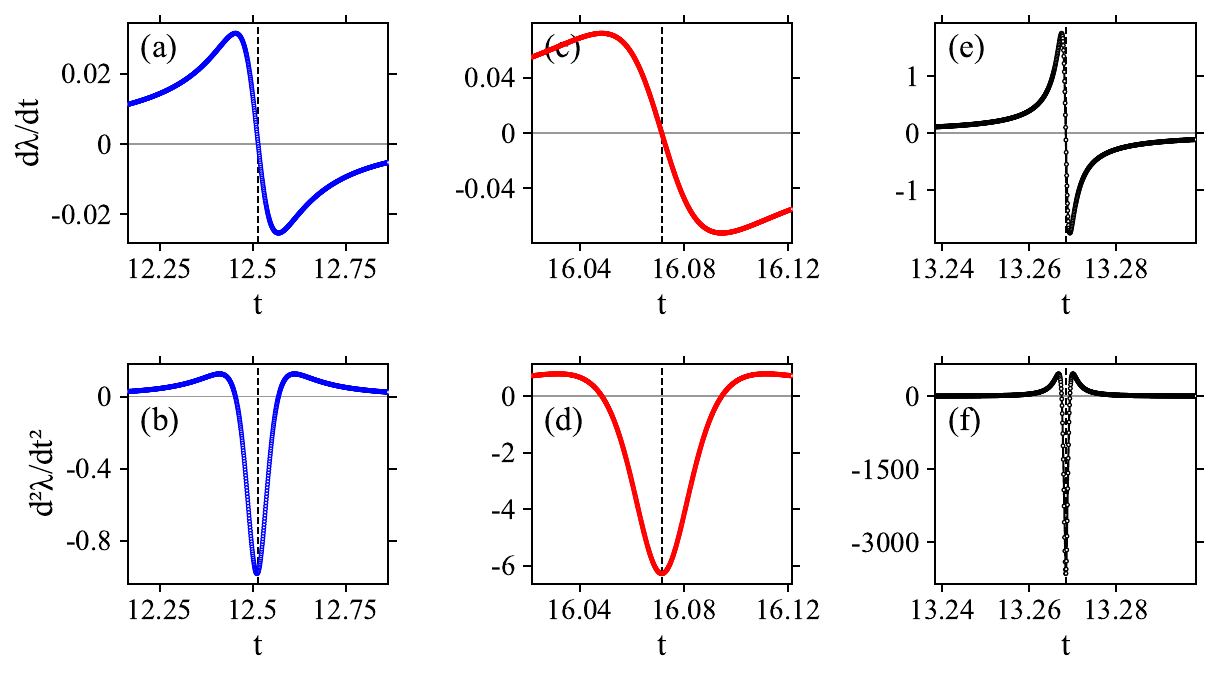}
  \caption{First and second order time derivatives of $\lambda(t)$ at $j=300$ for the $\Sigma$-, $\chi$-, and $\Lambda$-quenches. Panels (a,c,e): $d\lambda/dt$; panels (b,d,f): $d^2\lambda/dt^2$. Dashed lines mark $t_c$ and grey lines mark zero. A sign-changing feature appears in the first derivative at $t_c$ for the $\Sigma$- and $\Lambda$-quenches, while for the $\chi$-quench it emerges only in the second derivative, consistent with first- and second-order DQPTs, respectively.}
    \label{D}
\end{figure*}

\section{Complex-Time Fisher Zeros and the Order of DQPT's}
\label{FZ}
%%%%%%%%%%%%%%%%%%%

To investigate the origin of the DQPTs identified in the previous sections, we analytically continue the real-time Loschmidt amplitude into the complex-time plane by introducing the variable $z=t+i\tau \in \mathbb{C}$ \cite{PhysRevLett.110.135704,heyl2018dynamical}. The generalized Loschmidt amplitude is defined as

\begin{equation}
g(z)=\langle \psi_i|e^{-iHz}|\psi_i\rangle, \qquad z\in\mathbb{C},
\end{equation}
where $|\psi_i\rangle$ denotes the initial ground state and $H$ is the post-quench Hamiltonian. Expanding in the complete eigenbasis $\{|E_\nu\rangle\}$ of $H$ yields

\begin{equation}
g(z)= \sum_\nu
\left| \langle E_\nu|\psi_i\rangle \right|^2 e^{-iE_\nu z},
\label{eq:gz}
\end{equation}
where $E_\nu$ are the corresponding eigenenergies. For a finite system, Eq.~(\ref{eq:gz}) is a finite sum of analytic exponential functions and therefore defines an entire function of the complex variable $z$. Consequently, the Weierstrass factorization theorem can be applied to express the Loschmidt amplitude in terms of its complex zeros \cite{conway2012functions},

\begin{equation}
g(z) = e^{\mu(z)} \prod_j \left(z_j-z\right),
\label{eq:weierstrass}
\end{equation}
where $z_j$ are the Fisher zeros of the Loschmidt amplitude and $\mu(z)$ is an entire analytic function that contains the zero-free part of the amplitude and ensures convergence of the factorization. This representation explicitly separates the analytic and singular contributions. Since $\mu(z)$ is analytic everywhere in the complex plane, it cannot generate nonanalytic behavior. The singular contribution to the dynamical free-energy density (or rate function) is therefore determined entirely by the Fisher zeros,

\begin{equation}
g_s(z) = -\frac{1}{N}
\sum_j \ln\!\left(z_j-z\right),
\label{eq:gs}
\end{equation}

where $N$ denotes the system size. For finite systems, the Fisher zeros appear as isolated points in the complex-time plane. As the system size increases, the density of zeros grows and the discrete roots progressively condense into continuous lines or bounded regions. In the thermodynamic limit, nonanalyticities in the Loschmidt rate function emerge when these accumulations of Fisher zeros intersect the real-time axis. The corresponding crossing times coincide with the critical times of the DQPTs. Consequently, the geometry and accumulation pattern of Fisher zeros provide a microscopic characterization of dynamical criticality and allow the order and scaling properties of DQPTs to be analyzed in direct analogy with equilibrium phase transitions.

To complement the analysis of the Loschmidt return rate, we investigate the Fisher zeros of the analytically continued Loschmidt amplitude in the complex-time plane. The Fisher zeros, $z_k$, are defined by the condition $g(z_k)=0,$ and provide a complementary characterization of dynamical critical behaviour. In the thermodynamic limit, an accumulation of Fisher zeros approaching the real-time axis can generate non-analyticities in the Loschmidt return rate, thereby signaling a DQPT. Figure~\ref{FS} presents the Fisher-zero distributions for the three quench protocols considered in this work at $j=300$: (a) $\Sigma_f=1.0$, (b) $\chi_f=1.367$, and (c) $\Lambda_f=1.0$.

For the $\Sigma$-driven quench [Fig.~\ref{FS}(a)], the Fisher zeros form a complex-time structure associated with the dynamical critical time $t_c\simeq12.52$, consistent with the singularity identified in the Loschmidt return-rate analysis [Fig.~\ref{R2}]. The proximity of the zeros to the real-time axis near this time provides complementary evidence for the corresponding DQPT. For the $\chi$-driven quench [Fig.~\ref{FS}(b)], the Fisher-zero distribution is associated with the critical time $t_c\simeq16.06$, indicated by the vertical reference line [Fig.~\ref{R6}]. Similarly, the $\Lambda$-driven quench [Fig.~\ref{FS}(c)] exhibits a Fisher-zero structure associated with the critical time $t_c\simeq13.26$ [Fig.~\ref{R4}]. The distinct location of the Fisher-zero distributions reflect differences in the post-quench excitation spectra and in the overlap structure between the initial state and the eigenstates of the final Hamiltonian. Consequently, quenches along the $\Sigma$, $\chi$, and $\Lambda$ directions generate distinct complex-time signatures of dynamical criticality within the same many-body model.

The Fisher-zero distributions provide a complex-time perspective on the DQPTs identified from the return-rate function. However, the present calculations are performed for a system size ($j=300$), and the observed proximity of the zeros to the real-time axis does not establish a thermodynamic-limit pinch by itself. Demonstrating such behaviour requires a systematic finite-size analysis of the relevant Fisher-zero trajectories and, in particular, of the scaling of their imaginary parts with system size. Moreover, the order of a DQPT cannot be inferred solely from the visual arrangement of Fisher zeros. In the present work, the classification of the $\Sigma$- and $\Lambda$-driven transitions as first-order DQPTs and the $\chi$-driven transition as a second-order DQPT is based on the singular behaviour of the Loschmidt return rate and its derivatives, with Fisher-zero analysis serving as complementary evidence. Taken together, the real-time and complex-time analyses provide a consistent picture of dynamical criticality in the extended Agassi model and support the channel-dependent scaling behaviour revealed by the dynamical critical exponents.

A closer inspection of the Fisher-zero distributions reveals qualitative differences among the three quench protocols. The $\Sigma$- and $\Lambda$-driven quenches [Figs.~\ref{FS}(a) and~\ref{FS}(c)] exhibit comparatively dense distributions of zeros near the real-time axis, whereas the $\chi$-driven quench [Fig.~\ref{FS}(b)] displays a more dispersed distribution. These differences suggest distinct singularity structures in the corresponding Loschmidt return-rate functions. However, the order of a DQPT cannot be established solely from the visual arrangement of Fisher zeros and must instead be assessed through the behavior of the return-rate function and its derivatives.

To characterize the order of the dynamical transitions, we analyze the first and second time derivatives of the return-rate function $\lambda(t)$ for $j=300$, as shown in Fig.~\ref{D}. For the $\Sigma$-driven quench, Figs.~\ref{D}(a) and~\ref{D}(b) display pronounced structures in $d\lambda/dt$ and $d^2\lambda/dt^2$, respectively, near the critical time $t_c\simeq12.52$. The corresponding results for the $\chi$-driven quench are shown in Figs.~\ref{D}(c) and~\ref{D}(d), around $t_c\simeq16.06$, while Figs.~\ref{D}(e) and~\ref{D}(f) show the derivatives for the $\Lambda$-driven quench near $t_c\simeq13.26$. The derivative profiles support different classifications for these transitions. For the $\Sigma$- and $\Lambda$-driven quenches, the pronounced structures in the first derivative are consistent with first-order DQPTs, characterized by a nonanalyticity in $d\lambda/dt$ in the thermodynamic limit. In contrast, for the $\chi$-driven quench, the first derivative remains continuous across the critical time, while the dominant singular structure appears in the second derivative, consistent with a second-order DQPT. The Fisher-zero distributions provide complementary information about these transitions, with the relevant zeros occurring at the critical times identified from the real-time dynamics. Taken together, the derivative profiles and Fisher-zero distributions provide complementary signatures of the dynamical critical behavior in the extended Agassi model. The finite-size evolution of the derivative structures and the associated rounding effects are discussed in Appendix~\ref{FINITE SIZE DERIVATIVES}. Although the singularities remain rounded at finite $j$, these results are consistent with the emergence of nonanalytic behavior in the thermodynamic limit.

\section{Conclusion}
\label{conclusion}

We have investigated equilibrium quantum phase transitions and non-equilibrium DQPTs in the extended Agassi model. Following sudden quenches of the same-level pairing $\Sigma$, the monopole--monopole  $\chi$, and the cross-level pairing  $\Lambda$ strengths, we analyzed the Loschmidt echo and the associated return-rate function for system size $j=300$. The emergence of pronounced cusp-like structures in the return-rate function, together with their evolution with system size, provides finite-size signatures of DQPTs in all three parameter channels. To establish the equilibrium critical properties using finite-size scaling analysis of the order parameter $\mathcal{OP}_{J^2}$. The continuous symmetric-to-Hartree--Fock (HF) transition driven by $\chi$ exhibits mean-field critical behavior, with the order-parameter exponent approaching $\beta=1/2$ and reaching $\beta_\chi=0.5267\pm0.0035$ for $j=300$. In contrast, the $\Sigma$- and $\Lambda$-driven transitions are first order, characterized by discontinuous changes in the equilibrium order parameter rather than conventional power-law critical scaling.

We further quantified the non-equilibrium critical behavior by extracting dynamical singularity exponents from the local power-law scaling of the Loschmidt return-rate function near the critical times. For $j=300$, we obtained
$\alpha_\chi = 0.6225\pm0.0285, \alpha_\Sigma = 1.5373\pm0.1243, \alpha_\Lambda = 0.3191\pm0.0124$.
These estimates are near $ 2/3$, $3/2$, and $1/3$, respectively, suggesting distinct dynamical scaling behaviors for the three quench protocols. In particular, the $\Sigma$- and $\Lambda$-driven quenches exhibit substantially different dynamical exponents despite being associated with first-order equilibrium transitions. This contrast indicates that the local singularity structure of a DQPT is not determined solely by the equilibrium order of the transition, but also depends on the parameter channels and the post-quench excitation spectrum.  We further investigated the Fisher zeros of the analytically continued Loschmidt amplitude in the complex-time plane and examined the first and second derivatives of the return-rate function. The Fisher-zero distributions provide a complementary characterization of the dynamical critical times, while the derivative profiles support the classification of the $\Sigma$- and $\Lambda$-driven transitions as first-order DQPTs and the $\chi$-driven transition as a second-order DQPT. Although the finite-size results exhibit pronounced singularity precursors, establishing their precise thermodynamic-limit behaviour requires controlled finite-size extrapolations of the relevant zero trajectories and derivative structures.

Overall, the extended Agassi model provides a unified algebraic framework for investigating the interplay between equilibrium criticality and non-equilibrium dynamical criticality in a fermionic many-body system. The coexistence of continuous and discontinuous equilibrium transitions, together with the distinct dynamical scaling observed across the three quench parameter channels, makes the model a useful platform for exploring the emergence of dynamical universality. Future studies should pursue systematic finite-size scaling with quenched parameters and investigations of the entanglement dynamics \cite{PhysRevResearch.3.L042027} and universality. Further extensions to excited-state quantum phase transitions \cite{PhysRevA.78.060102, PhysRevA.100.062113} and out-of-time-order correlators \cite{CAPRIO20081106} could provide additional insight into the relationship between spectral structure, quantum correlations, and non-equilibrium critical behaviour. These directions may help establish whether the observed fractional exponents represent genuine asymptotic scaling laws and clarify the broader universality properties of algebraic many-body systems.

\section*{Author Contributions}

Ritesh Kumar Palai and Rajiv Kumar contributed equally to this work and share first authorship.

\section*{Acknowledgement}
RKP acknowledge the Central University of South Bihar (CUSB) Gaya, Bihar,  IIT (BHU) Varanasi, Uttar Pradesh and, Department of Science and Technology (DST), Government of India.
RK also acknowledge financial support from the Council of Scientific and Industrial Research (CSIR) fellowship, New Delhi, India. 

\section*{DATA AVAILABILITY}
\label{DATA}
The data supporting the findings of this article are not publicly available. The data are available from the authors upon reasonable request.

\appendix
\section{Exact diagonalization in the seniority-zero pair basis}
\label{app:hamiltonian}

The extended Agassi Hamiltonian [Eq.~(\ref{eq.Ham})], and the associated fermionic operators mentioned in Eq.~(\ref{eq.OP1}) obey the \(\mathfrak{so}(5)\) Lie algebra.
Following Ref.~\cite{PhysRevC.95.064306, GarciaRamos2019}, a complete (though non‑orthogonal) basis
is given by the states. The physical subspace with zero seniority restricts the system to $n=0$. The exact diagonalization is performed directly in this collective subspace by constructing the orthonormal basis $\ket{n_1, n_{-1}, n_0}$.  This procedure completely bypasses the overlap matrix inherent to non-orthogonal basis, enabling highly stable, sparse diagonalization even for large system sizes.
\begin{equation}
|n_-,n_+,n_0\rangle =
(A_{-1}^\dagger)^{n_-}(A_1^\dagger)^{n_+}(A_0^\dagger)^{n_0}|0\rangle,
\label{A3}
\end{equation}
where \(n_-\) and \(n_+\) are the numbers of pairs in the lower and upper levels, respectively, and \(n_0\) is the number of pairs \(A_0^\dagger\). For the half‑filled case \(N=2j\), one has
\begin{equation}
n_- + n_+ + n_0 = j.
\label{A4}
\end{equation}

\begin{figure}[htb]
    \centering
    \includegraphics[width=0.49\linewidth, height=0.5\linewidth]{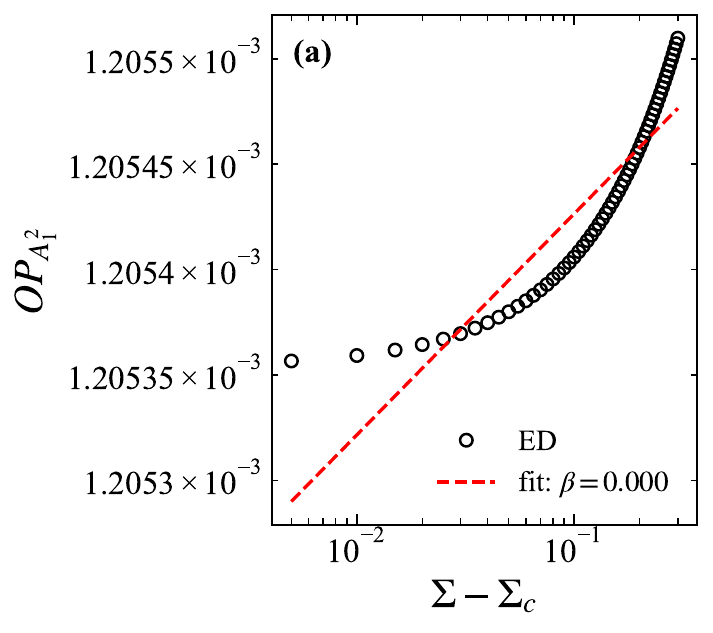}
    \includegraphics[width=0.49\linewidth, height=0.5\linewidth]{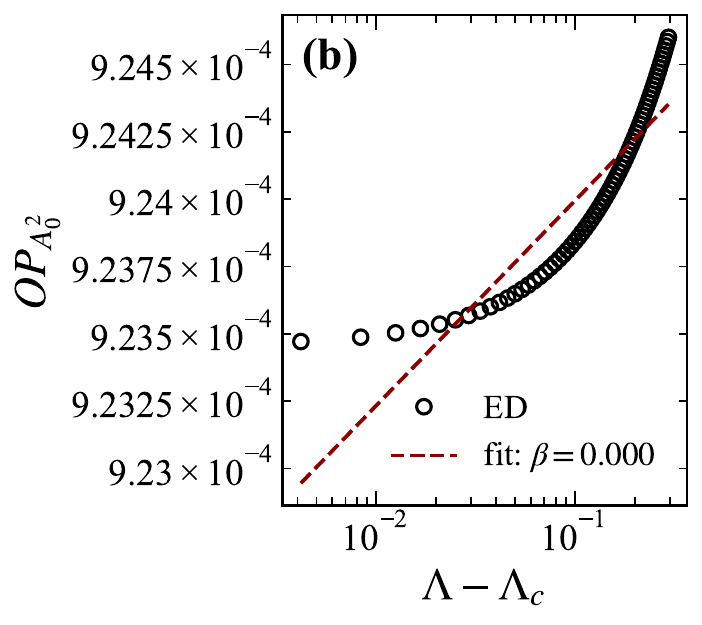}
  \caption{The log--log behavior of the order parameters for $j=300$ from exact diagonalization (ED) in the seniority-zero basis with $\chi=1.5$. 
(a) $\Sigma$-driven: $OP_{A_1^2}$ vs $\Sigma-\Sigma_c$, with $\Sigma_c=1.55500$ from the gap minimum. Open circles are ED data; the red dashed line is a power-law fit over $\Sigma_c<\Sigma<\Sigma_c+0.3$, giving $\beta=0.0000\pm0.0000$. 
(b) $\Lambda$-driven: $OP_{A_0^2}$ vs $\Lambda-\Lambda_c$, with $\Lambda_c=1.09500$; same procedure yields $\beta=0.0004\pm0.0000$.} \label{AP0}
\end{figure} 

\begin{figure}
    \centering
    \includegraphics[width=0.9\linewidth, height=0.8\linewidth]{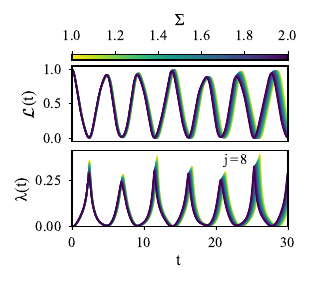}
    \includegraphics[width=0.84\linewidth, height=0.75\linewidth]{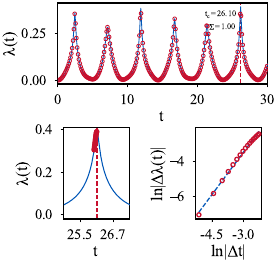}
  \caption{Same caption as Fig.~\ref{sig_los} and Fig.~\ref{R2} but for $j=8$, with the strongest DQPT signature at $\Sigma_f=1.0$, $t_c=26.10$, $\alpha_\Sigma=1.613\pm0.028$.}
    \label{AP5}
\end{figure}

\begin{figure}
    \centering
    \includegraphics[width=0.92\linewidth, height=0.8\linewidth]{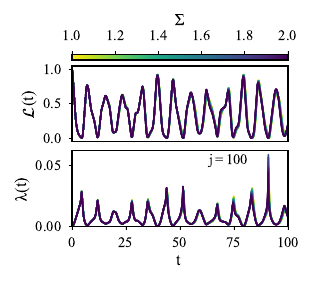}
    \includegraphics[width=0.84\linewidth, height=0.75\linewidth]{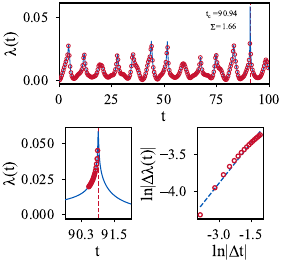}
   \caption{Same caption as Fig.~\ref{sig_los} and Fig.~\ref{R2} but for $j=100$, with the strongest DQPT signature at $\Sigma_f=1.655$, $t_c=90.94$, $\alpha_\Sigma=0.354\pm0.013$.}
    \label{AP6}
\end{figure}

\begin{figure}
    \centering
    \includegraphics[width=0.9\linewidth, height=0.8\linewidth]{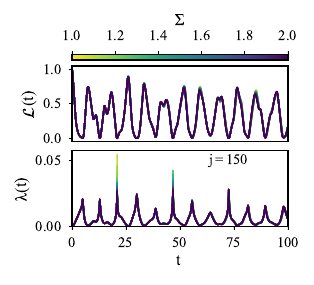}
    \includegraphics[width=0.84\linewidth, height=0.75\linewidth]{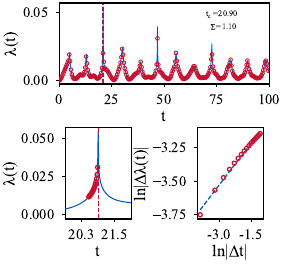}
    \caption{Same caption as Fig.~\ref{sig_los} and Fig.~\ref{R2} but for $j=150$, with the strongest DQPT signature at $\Sigma_f=1.103$, $t_c=20.90$, $\alpha_\Sigma=0.204\pm0.005$.}
    \label{AP7}
\end{figure}

\begin{figure}
    \centering
    \includegraphics[width=0.9\linewidth, height=0.8\linewidth]{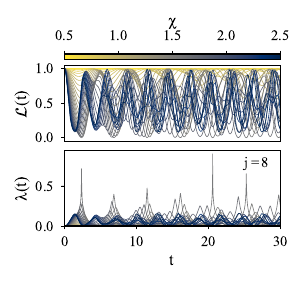}
    \includegraphics[width=0.84\linewidth, height=0.75\linewidth]{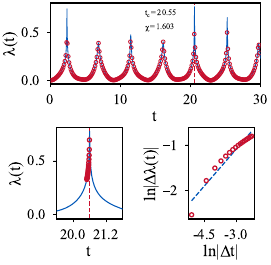}
    \caption{Same caption as Fig.~\ref{R5} and Fig.~\ref{R6} but for $j=8$, with the strongest DQPT signature at $\chi_f=1.603$, $t_c=20.55$, and $\alpha_\chi=0.536\pm0.036$.}
    \label{AP8}
\end{figure}

\begin{figure}
    \centering
    \includegraphics[width=0.9\linewidth, height=0.8\linewidth]{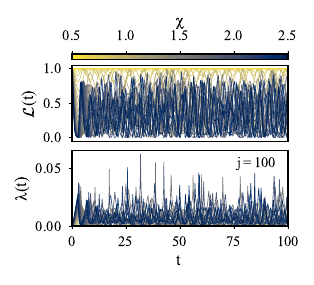}
    \includegraphics[width=0.84\linewidth, height=0.75\linewidth]{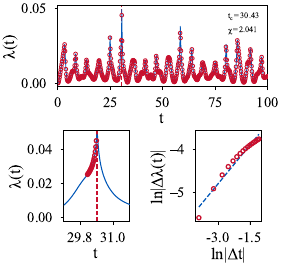}
    \caption{Same caption as Fig.~\ref{R5} and Fig.~\ref{R6} but for $j=100$, with the strongest DQPT signature at $\chi_f=2.041$, $t_c=30.43$, and $\alpha_\chi=0.600\pm0.031$.}
    \label{AP9}
\end{figure}

\begin{figure}
    \centering
    \includegraphics[width=0.9\linewidth, height=0.8\linewidth]{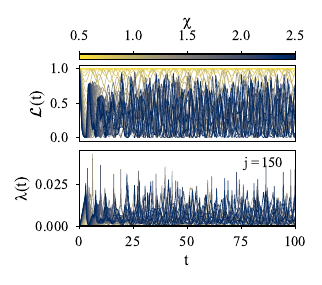}
    \includegraphics[width=0.84\linewidth, height=0.75\linewidth]{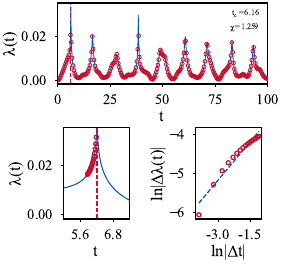}
     \caption{Same caption as Fig.~\ref{R5} and Fig.~\ref{R6} but for $j=150$, with the strongest DQPT signature at $\chi_f=1.259$, $t_c=6.16$, and $\alpha_\chi=0.647\pm0.037$.}
     \label{AP10}
\end{figure}

\begin{figure}
    \centering
    \includegraphics[width=0.9\linewidth, height=0.8\linewidth]{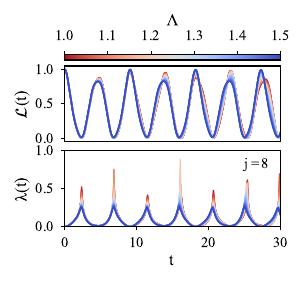}
    \includegraphics[width=0.84\linewidth, height=0.75\linewidth]{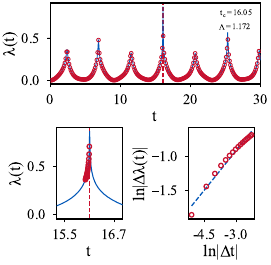}
   \caption{Same caption as Fig.~\ref{R3} and  Fig.~\ref{R4} but for $j=8$, with the strongest DQPT signature at $\Lambda_f=1.172$, $t_c=16.05$, and $\alpha_\Lambda=0.382\pm0.016$.}
    \label{AP2}
\end{figure}

\begin{figure}
    \includegraphics[width=0.9\linewidth, height=0.8\linewidth]{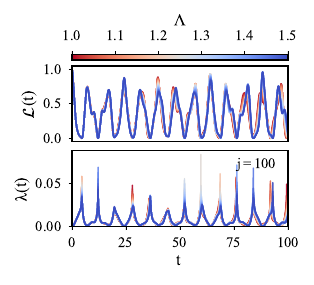}
    \includegraphics[width=0.81\linewidth, height=0.75\linewidth]{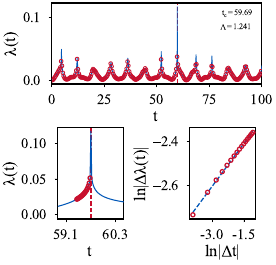}
    \caption{Same caption as Fig.~\ref{R3} and Fig.~\ref{R4} but for $j=100$, with the strongest DQPT signature at $\Lambda_f=1.241$, $t_c=59.69$, and $\alpha_\Lambda=0.130\pm0.001$.}
    \label{AP3}
\end{figure}

\begin{figure}
    \includegraphics[width=0.9\linewidth, height=0.8\linewidth]{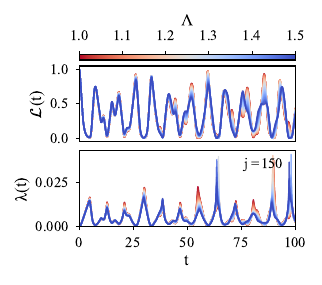}
    \includegraphics[width=0.84\linewidth, height=0.75\linewidth]{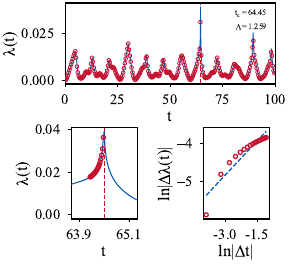}
   \caption{Same caption as Fig.~\ref{R3} and Fig.~\ref{R4} but for $j=150$, with the strongest DQPT signature at $\Lambda_f=1.259$, $t_c=64.45$, and $\alpha_\Lambda=0.591\pm0.058$.}
    \label{AP4}
\end{figure}

The action of the generators on these states is given that operators $J$ and $A$ are
\begin{align*}
J_{+}\ket{n_{-},n_{+},n_{0}}
&= n_{+}\ket{n_{-},n_{+}-1,n_{0}+1} \\ \nonumber
&+2n_{0}\ket{n_{-},n_{+}+1,n_{0}-1},
\\  \tag{A3}
J_{-}\ket{n_{-},n_{+},n_{0}}
&= n_{-}\ket{n_{-}-1,n_{+},n_{0}+1}   \\ \nonumber
&+2n_{0}\ket{n_{-}+1,n_{+},n_{0}-1}, \\
J_{0}\ket{n_{-},n_{+},n_{0}}
&=(n_{+}-n_{-})\ket{n_{-},n_{+},n_{0}},\\
A_{+}^{\dagger}\ket{n_{-},n_{+},n_{0}}
&=\ket{n_{-},n_{+}+1,n_{0}},\\
A_{+}\ket{n_{-},n_{+},n_{0}}
&=n_{+}(j-n_{+}+1-n_{0})
\ket{n_{-},n_{+}-1,n_{0}} \\ \nonumber  
&-n_{0}(n_{0}-1)
\ket{n_{-},n_{+}+1,n_{0}-2},\\
A_{-}^{\dagger}\ket{n_{-},n_{+},n_{0}}
&=\ket{n_{-}+1,n_{+},n_{0}},
\\
A_{-}\ket{n_{-},n_{+},n_{0}}
&=n_{-}(j-n_{-}+1-n_{0})
\ket{n_{-}-1,n_{+},n_{0}} \\ \nonumber
&-n_{0}(n_{0}-1)
\ket{n_{-}+1,n_{+},n_{0}-2},\\
A_{0}^{\dagger}\ket{n_{-},n_{+},n_{0}}
&=\ket{n_{-},n_{+},n_{0}+1}, \\
\end{align*}

\begin{align*}
A_{0}\ket{n_{-},n_{+},n_{0}}
&=n_{0}(2j-2n_{0}-2n_{-}-2n_{+}+1)
\ket{n_{-},n_{+},n_{0}-1} \\ \nonumber
&-n_{-}n_{+}
\ket{n_{-}-1,n_{+}-1,n_{0}+1}, \\
N\ket{n_{-},n_{+},n_{0}}
&=2(n_{-}+n_{+}+n_{0})
\ket{n_{-},n_{+},n_{0}}.
\end{align*}

with the corresponding expressions for \(A_{-1}\), \(A_0\), and their Hermitian conjugates.

\section{First Order Equilibrium Critical Exponents}
The phases changes about $\Sigma$ and $\Lambda$ refer to a first order phase transition. We show the order parameter in Figure~\ref{AP0} (a) for $\Sigma$ and (b) for $\Lambda$, displaying the log-log order parameter versus the reduced control parameter for $j=300$ in both driving protocols. power-law fits yield vanishing effective exponents $\beta=0.0000$ for $\Sigma$ and $0.0004$ for $\Lambda$, signaling a first-order transition rather than a continuous one.

\section{Finite-Size Evolution of the \texorpdfstring{$\Sigma$}{Sigma}-Driven DQPT}
\label{AA2}
The Loschmidt echo $\mathcal{L}(t)$ and the return rate function $\lambda(t)$ for $j=8$, $100$, and $150$ are shown in Figs.~\ref{AP5} to \ref{AP7}, illustrating the evolution of finite-size DQPT precursors with increasing system size. For finite $j$, the Loschmidt echo $\mathcal{L}(t)$ exhibits pronounced revivals, while the corresponding return-rate function $\lambda(t)$ displays rounded extrema that progressively sharpen as $j$ increases. The singularity precursors occur at $(\Sigma_f,t_c,\alpha)=(1.000,26.099,1.613\pm0.028)$, $(1.655,90.938,0.354\pm0.013)$, and $(1.103,20.904,0.204\pm0.005)$ for the system size $j=8, 100$, and $150$, respectively (see Table~\ref{tab:dqpt_exponents} for dynamical exponents). The quench parameter $\Sigma_f$ varies with system size, these results do not constitute a controlled finite-size scaling sequence, and the extracted exponents should therefore be interpreted as finite-size estimates rather than asymptotic values. Nevertheless, systematic sharpening of the return-rate structures and the persistence of coherent revivals over a finite range of $\Sigma_f$ support the existence of a DQPT in the thermodynamic limit and reflect the collective nature of the extended Agassi model.

\section{ Finite-Size Evolution Following a \texorpdfstring{$\chi$}{Chi} Quench}
\label{AA3}
The Loschmidt echo $\mathcal{L}(t)$ and the return rate function $\lambda(t)$ for $j=8$, $100$ and $150$ are shown in Figs.~\ref{AP8} to \ref{AP10}, illustrating the evolution of finite-size DQPT precursors under the same $\chi$-driven quench protocol. For finite $j$, the Loschmidt echo $\mathcal{L}(t)$ exhibits pronounced oscillations and revivals, while the corresponding return-rate function $\lambda(t)$ displays rounded extrema. As the size of the system increases, the $\mathcal{L}(t)$ and $\lambda(t)$ increases, leading to progressively sharper and more localized structures. The singularity precursors occur at $(\chi_f,t_c,\alpha)=(1.603,20.554,0.536\pm0.036)$, $(2.041,30.426,0.600\pm0.031)$, and $(1.259,6.161,0.647\pm0.037)$ for $j=8$, $100$, and $150$, respectively (see Table~\ref{tab:dqpt_exponents} for dyanmical exponents). The quench parameter $\chi_f$ varies with system size, the extracted exponents should be regarded as finite-size estimates. Nevertheless, systematic sharpening of the return-rate structures and the persistence of coherent revivals over a finite interval of $\chi_f$ provide clear evidence for the emergence of a DQPT in the thermodynamic limit.

\section{Finite-Size Evolution Following a \texorpdfstring{$\Lambda$}{Lambda} Quench}
\label{AA4}
The Loschmidt echo $\mathcal{L}(t)$ and the return rate function $\lambda(t)$ for $j=8$, $100$ and $150$ are shown in Figs.~\ref{AP2} - \ref{AP4}, illustrating the evolution of the finite-size DQPT signature with increasing system size. For finite $j$, the Loschmidt echo $\mathcal{L}(t)$ exhibits pronounced revivals, while the corresponding return-rate function $\lambda(t)$ displays rounded extrema that progressively sharpen as $j$ increases. The strongest singularity precursors occur at $(\Lambda_f,t_c,\alpha)=(1.172,16.053,0.382\pm0.016)$, $(1.241,59.692,0.130\pm0.001)$, and $(1.259,64.453,0.591\pm0.058)$ for $j=8$, $100$, and $150$, respectively (see Table~\ref{tab:dqpt_exponents} for dynamical exponents). The quench parameter $\Lambda_f$ varies with system size, these results do not constitute a controlled finite-size scaling sequence, and the extracted exponents should therefore be interpreted as finite-size estimates rather than asymptotic values. Nevertheless, systematic sharpening of the return-rate structures and the persistence of coherent revivals across a finite interval of $\Lambda_f$ highlight the collective nature of the extended Agassi model and support the existence of a DQPT in the thermodynamic limit.

Overall, the finite-size results presented in Figs.~\ref{AP5}-\ref{AP4} demonstrate a systematic evolution of the Loschmidt dynamics with increasing system size. For all three quench protocols, the Loschmidt echo exhibits pronounced revivals, and the corresponding return-rate function develops increasingly sharp and localized extrema as $j$ increases. These features constitute finite-size precursors of the non-analytic cusps observed for $j=300$ in the main text. Although the extracted critical times and dynamical exponents shows, the strongest singularities generally become more pronounced with increasing system size, indicating the gradual emergence of dynamical critical behaviour in the thermodynamic limit. A quantitative finite-size scaling analysis is complicated by the fact that the quench parameters $(\chi_f,\Sigma_f,\Lambda_f)$ associated with the DQPT precursor vary with $j$, preventing a controlled extrapolation of the dynamical exponents. Nevertheless, the results $j=300$ consistently represent the sharpest singular structures and therefore provide the most reliable estimates of asymptotic dynamical critical behaviour. Taken together, the finite-size analysis support the existence of DQPTs in all three parameter channels and establish the channels-dependent exponents $\alpha_\chi$, $\alpha_\Sigma$, and $\alpha_\Lambda$ obtained for $j=300$ as representative signatures of the underlying non-equilibrium critical dynamics of the extended Agassi model.

\begin{figure*}
    \centering
    \includegraphics[width=0.3\linewidth, height=0.28\linewidth]{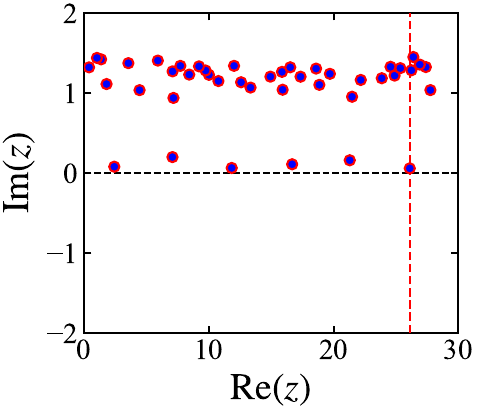}
    \includegraphics[width=0.3\linewidth, height=0.28\linewidth]{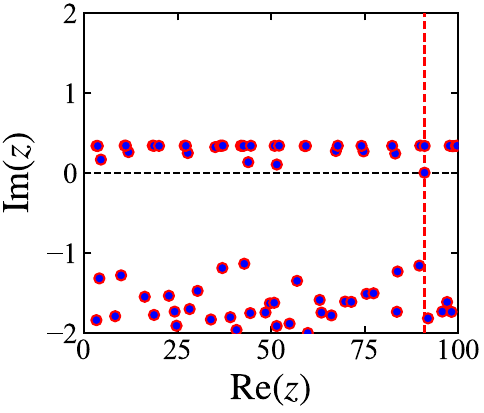}
    \includegraphics[width=0.3\linewidth, height=0.28\linewidth]{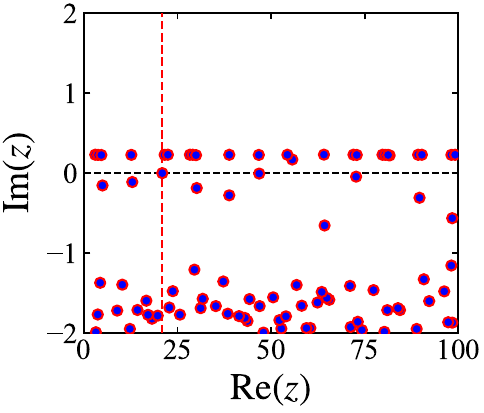}
    \caption{Same caption as Fig.~\ref{FS}(a) but for $j=8$, $100$, and $150$, with $\Sigma_f=1.0$, $t_c\simeq26.10$; $\Sigma_f=1.655$, $t_c\simeq90.94$; and $\Sigma_f=1.103$, $t_c\simeq20.90$, respectively.}
    \label{FS1}
\end{figure*}

\begin{figure*}
    \centering
    \includegraphics[width=0.3\linewidth, height=0.28\linewidth]{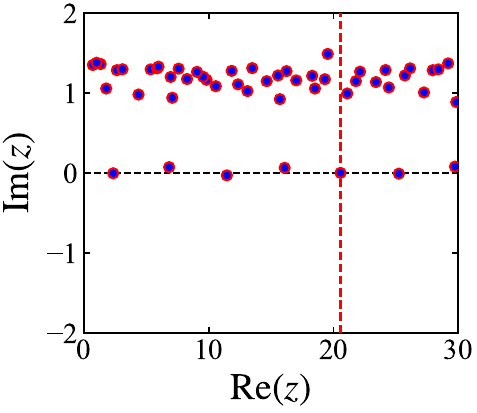}
    \includegraphics[width=0.3\linewidth, height=0.28\linewidth]{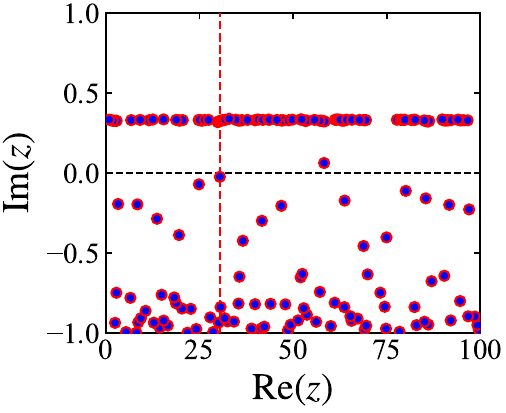}
    \includegraphics[width=0.3\linewidth, height=0.28\linewidth]{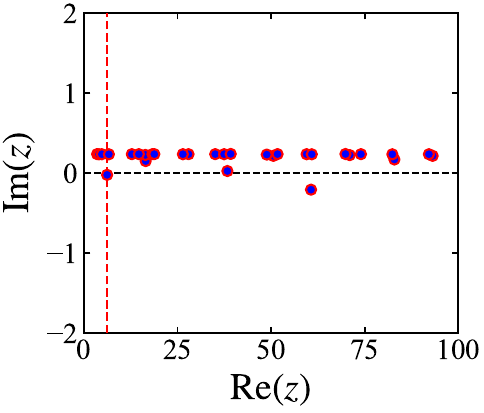}
    \caption{Same caption as Fig.~\ref{FS}(b) but for $j=8$, $100$, and $150$, with $\chi_f=1.603$, $t_c\simeq20.55$; $\chi_f=2.041$, $t_c\simeq30.43$; and $\chi_f=1.259$, $t_c\simeq6.16$, respectively.}
    \label{FS2}
\end{figure*}

\begin{figure*}
    \centering
    \includegraphics[width=0.3\linewidth, height=0.28\linewidth]{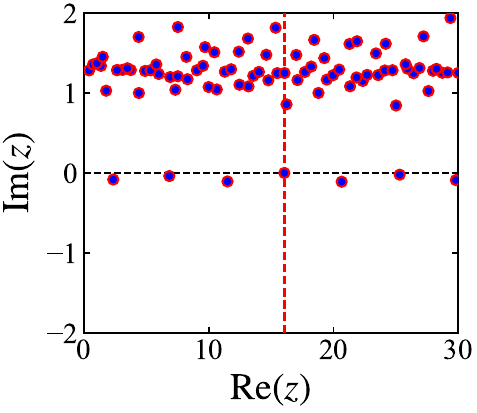}
    \includegraphics[width=0.3\linewidth, height=0.28\linewidth]{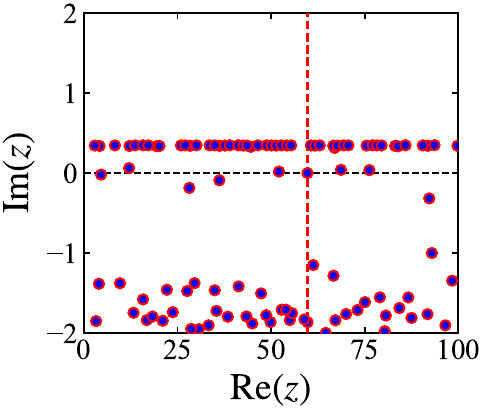}
    \includegraphics[width=0.3\linewidth, height=0.28\linewidth]{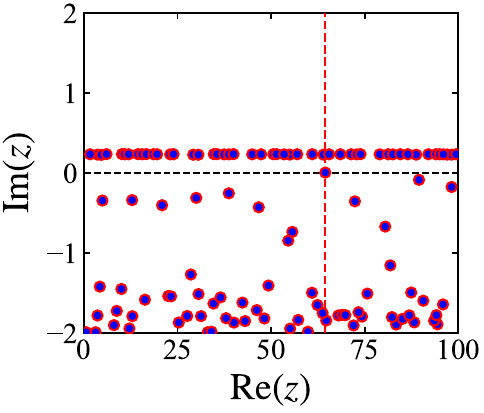}
    \caption{Same caption as Fig.~\ref{FS}(c)  but for $j=8$, $100$, and $150$, with $\Lambda_f=1.172$, $t_c\simeq16.05$; $\Lambda_f=1.241$, $t_c\simeq59.69$; and $\Lambda_f=1.259$, $t_c\simeq64.45$, respectively.}
    \label{FS3}
\end{figure*}

\begin{figure*}
    \centering
    \includegraphics[width=0.9\linewidth, height=0.5\linewidth]{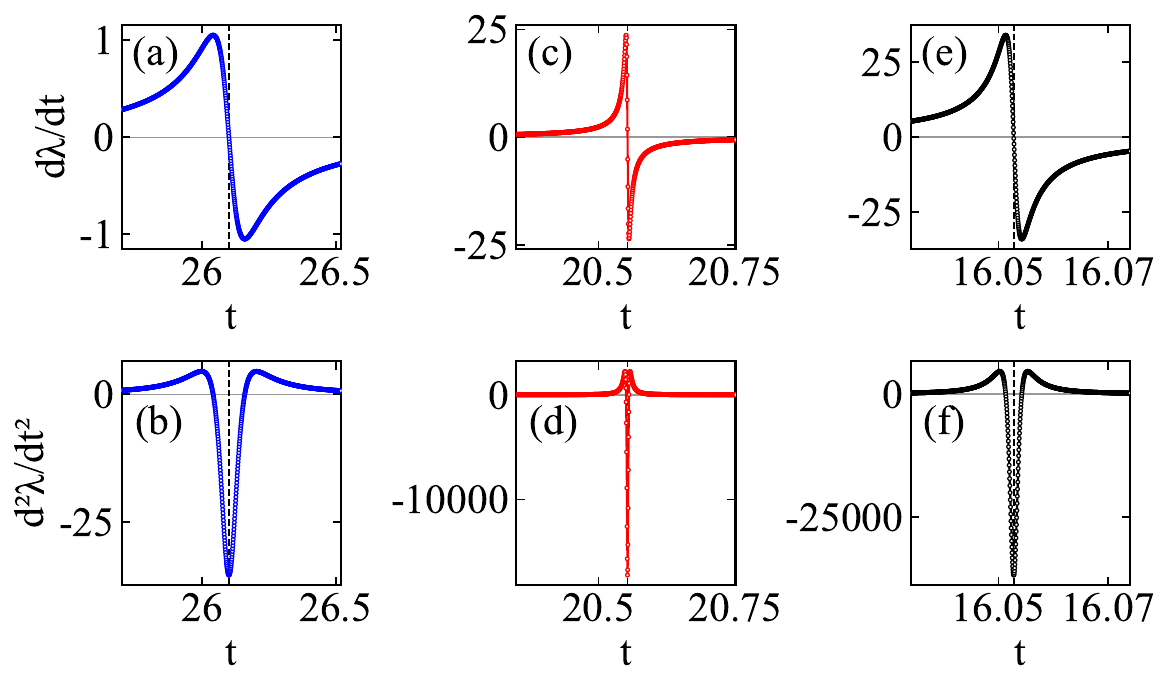}
    \caption{Time evolution of the first and second derivatives of the rate function, $d\lambda/dt$ and $d^2\lambda/dt^2$, for the $\Sigma$-, $\chi$-, and $\Lambda$-quenches at finite system size $j=8$.}
    \label{D1}
\end{figure*}

\begin{figure*}
    \centering
    \includegraphics[width=0.9\linewidth, height=0.5\linewidth]{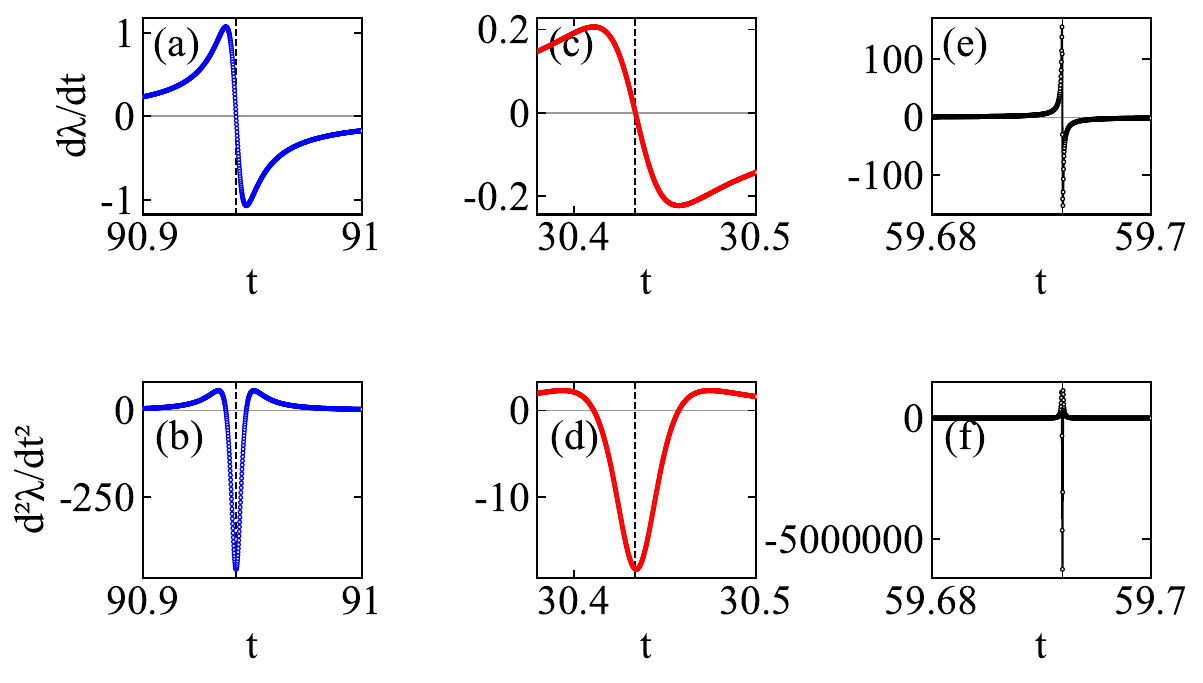}
    \caption{Time evolution of the first and second derivatives of the rate function, $d\lambda/dt$ and $d^2\lambda/dt^2$ for the $\Sigma$-, $\chi$-, and $\Lambda$-quenches at system size $j=100$. }
    \label{D2}
\end{figure*}

\begin{figure*}
    \centering
    \includegraphics[width=0.9\linewidth, height=0.5\linewidth]{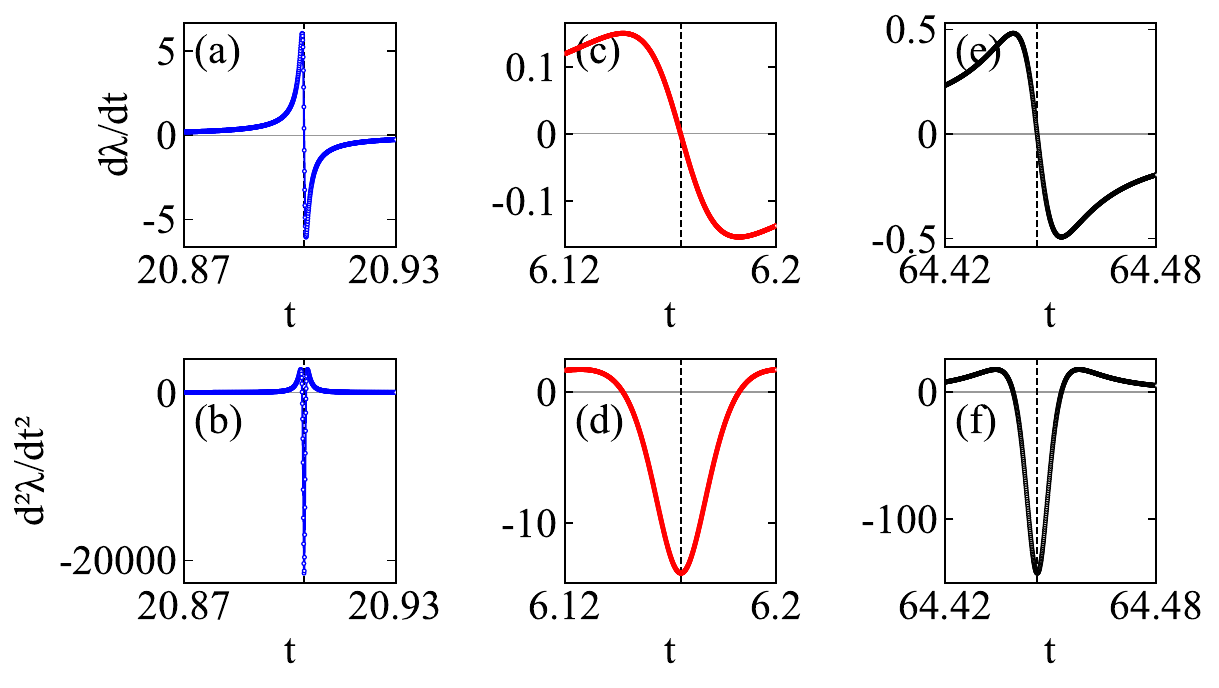}
    \caption{Time evolution of the first and second derivatives of the rate function, $d\lambda/dt$ and $d^2\lambda/dt^2$ for the $\Sigma$-, $\chi$-, and $\Lambda$-quenches at system size $j=150$.}
    \label{D3}
\end{figure*}

\section{Finite Size Effects Fisher zeros and  DQPT Order from Rate-Function Derivatives}
\label{FINITE SIZE DERIVATIVES}
We further examine the Fisher zeros in the complex-time plane for the parameters $\Sigma$, $\chi$, and $\Lambda$ for finite system sizes $j=8$, $100$, and $150$ shown in Figs. \ref{FS1} to \ref{FS3}. In all Figures, the zeros approach and intersect the real-time axis near the corresponding critical time, marked by the vertical red dashed line, providing a signature of the DQPT. With increasing $j$, the distribution of zeros becomes increasingly dense near the real axis, consistent with the non-analytic behaviour of the rate function.
The three parameter quenches exhibit distinct dynamical singularities. Combining the Fisher-zero structure with the corresponding behaviour of the rate function and its derivatives, we identify the $\Sigma$- and $\Lambda$-quenches as first-order DQPTs, whereas the $\chi$-quench exhibits a continuous, second-order-like singularity. The finite-size Fisher-zero distributions alone do not constitute a thermodynamic-limit classification; the above assignments refer to the singularity behaviour observed from the combined analyses.

The order of the DQPTs can be inferred from the non-analytic behaviour of the rate function and its time derivatives at the critical time \(t_c\). As shown in Figs.~\ref{D1}--\ref{D3}, the \(\Sigma\)-, \(\Lambda\)-, and \(\chi\)-quenches exhibit distinct signatures in the first and second derivatives of the rate function for finite system sizes \(j=8,100,\) and \(150\). For the \(\Sigma\)- and \(\Lambda\)-quenches (Figs.~\ref{D1} and \ref{D3}), the first derivative, \(d\lambda/dt\), exhibits a pronounced sharp variation near \(t_c\), accompanied by a discontinuous peak in the second derivative, \(d^2\lambda/dt^2\). These features are consistent with a first-order-like DQPT, characterized by a non-analyticity in the first derivative of the rate function. In contrast, for the \(\chi\)-quench (Fig.~\ref{D2}), the first derivative remains comparatively smooth near \(t_c\), whereas the second derivative develops a sharp, localized peak, suggesting a second-order-like dynamical phase transition. For finite \(j\), these features remain rounded rather than representing genuine discontinuities. Nevertheless, the sharpening of the second-derivative feature with increasing system size suggests that it may develop into a divergence or discontinuity in the thermodynamic limit. Thus, the observed behavior is consistent with a non-analyticity emerging primarily in the second derivative for the \(\chi\)-quench, whereas the \(\Sigma\)- and \(\Lambda\)-quenches exhibit signatures associated with a non-analyticity in the first derivative. Since the results are obtained for finite system sizes, these classifications should be regarded as indicative rather than conclusive. A systematic finite-size scaling analysis of the peak heights and widths is required to establish the order of the transitions in the thermodynamic limit.

\bibliography{ref}

\end{document}